\documentclass[aps, prd, twocolumn,showpacs, showkeys, preprintnumbers, nofootinbib, superscriptaddress]{revtex4-2}
\usepackage[sort&compress]{natbib}
\usepackage{multirow}
\usepackage{graphicx}
\usepackage{amssymb,amsbsy,amsmath,amsfonts}
\usepackage{slashed}
\usepackage[latin1]{inputenc}

\begin{document}
\title{$\phi(2170)$ decaying to $\phi\eta$ and $\phi\eta^\prime$}
\author{Brenda B. Malabarba}
\email{brenda@if.usp.br}
\affiliation{Universidade de Sao Paulo, Instituto de Fisica, C.P. 05389-970, Sao 
Paulo, Brazil.}

\author{K. P. Khemchandani}
\email{kanchan.khemchandani@unifesp.br}
\affiliation{Universidade Federal de Sao Paulo, C.P. 01302-907, Sao Paulo, Brazil.}

\author{A. Mart\'inez Torres}
\email{amartine@if.usp.br}
\affiliation{Universidade de Sao Paulo, Instituto de Fisica, C.P. 05389-970, Sao 
Paulo, Brazil.}

\begin{abstract}
In this work we calculate the decay widths of $\phi(2170)$ to $\phi\eta$ and $\phi\eta^\prime$ by considering $\phi(2170)$ as a $\phi K\bar K$ state, with $K\bar K$ clustering as $f_0(980)$. These decay widths have been recently determined by the BESIII, BaBar and Belle collaborations with the aim of unraveling the nature of $\phi(2170)$. By analyzing the data on the cross sections of the process $e^+e^-\to\phi\eta$, two solutions were found by the BESIII collaboration with the same $\chi^2$ per number of degrees of freedom (n.d.f), mass and width for $\phi(2170)$. However, different values for $\mathcal{B}^{\phi(2170)}_{\phi\eta}\Gamma^{\phi(2170)}_{e^+e^-}$ were obtained, where $\mathcal{B}^{\phi(2170)}_{\phi\eta}$ is the branching fraction of $\phi(2170)\to\phi\eta$ and $\Gamma^{\phi(2170)}_{e^+e^-}$ is the decay width for $\phi(2170)\to e^+e ^-$: $0.24^{+0.12}_{-0.07}$ eV (solution I) and $10.11^{+3.87}_{-3.13}$ eV (solution II). In case of $\mathcal{B}^{\phi(2170)}_{\phi\eta^\prime}\Gamma^{\phi(2170)}_{e^+e^-}$, a value of $7.1\pm 0.7\pm 0.7$ was determined by the BESIII collaboration from fits to the data on the cross section of $e^+e^-\to\phi\eta^\prime$. The Belle collaboration has also determined $\mathcal{B}^{\phi(2170)}_{\phi\eta}\Gamma^{\phi(2170)}_{e^+e^-}$ more recently, although the statistical significance related to the signal of $\phi(2170)$ in the data is low. We calculate the decay widths of $\phi(2170)$ to $\phi\eta$ and $\phi\eta^\prime$, and compare their ratio $\Gamma^{\phi(2170)}_{\phi \eta}/\Gamma^{\phi(2170)}_{\phi \eta^\prime}$ with the one determined by using the above mentioned experimental data. Considering the theoretical and experimental uncertainties, the lower limit of our theoretical result ($\simeq 2.6$) is close to the upper value obtained ($\simeq 2$) by using the solution II of BESIII for $\mathcal{B}^{\phi(2170)}_{\phi\eta}\Gamma^{\phi(2170)}_{e^+e^-}$, as well as to the upper limit found ($\simeq 2.8$) when considering the solutions III and IV of the Belle collaboration for $\mathcal{B}^{\phi(2170)}_{\phi\eta}\Gamma^{\phi(2170)}_{e^+e^-}$.
\end{abstract}



\maketitle
\date{\today}

\section{Introduction}
From the perspective of the Okubo-Zweig-Iizuka (OZI) rule~\cite{Okubo:1963fa,Iizuka:1966fk}, the $\phi(2170)$ meson~\cite{BaBar:2006gsq,Belle:2008kuo,BES:2007sqy}, as an excited state of $\phi(1020)$, is expected to have non suppressed decay modes to the $\phi\eta$ and $\phi\eta^\prime$ channels. Determining the decay width of $\phi(2170)$ to these channels can thus provide valuable information on the nature of $\phi(2170)$, in view of the broad spectrum of interpretations suggested since its discovery: a $s\bar s$ state, a tetraquark, a hybrid state, a bound $\Lambda\bar \Lambda$ state, a three meson state~\cite{Barnes:2002mu,Ding:2007pc,Pang:2019ttv,Li:2020xzs,Wang:2021gle,Ding:2006ya,Page:1998gz,Barnes:1995hc,Ma:2020bex,Wang:2006ri,Chen:2008ej,Deng:2010zzd,Chen:2018kuu,Ke:2018evd,Wang:2019nln,Liu:2020lpw,Coito:2009na,Drenska:2008gr,Agaev:2019coa,Zhao:2013ffn,Dong:2017rmg,MartinezTorres:2008gy}. 

With this motivation, the BESIII collaboration has studied the processes $e^+ e^-\to \phi\eta$~\cite{BESIII:2021bjn} and $e^+ e^-\to \phi \eta^\prime$~\cite{BESIII:2020gnc}, where a signal for $\phi(2170)$ was observed in both cases (with a statistical significance of $6.9\sigma$ in the first case and larger than $10\sigma$ in the second). By fitting the data, two solutions with the same $\chi^2$/n.d.f, mass and width for $\phi(2170)$, but different value for the product $\mathcal{B}^{\phi(2170)}_{\phi\eta}\Gamma^{\phi(2170)}_{e^+e^-}$, were found in Ref.~\cite{BESIII:2021bjn}: $0.24^{+0.12}_{-0.07}$ eV (solution I) and $10.11^{+3.87}_{-3.13}$ eV (solution II), respectively. The first solution is compatible with the result earlier obtained by the BaBar collaboration, due to the large uncertainty in the latter, $1.7\pm 0.7\pm 1.3$~\cite{BaBar:2007ceh}. Recently, the Belle collaboration~\cite{Belle:2022fhh} has also determined the value for $\mathcal{B}^{\phi(2170)}_{\phi\eta}\Gamma^{\phi(2170)}_{e^+e^-}$ from fits to the data on $e^+e^-\to \phi\eta$, finding four possible solutions with the same $\chi^2$/n.d.f: $0.09\pm 0.05$ (solution I), $0.06\pm 0.02$ (solution II), $16.7\pm 1.2$ (solution III), $17.0\pm 1.2$ (solution IV). The results (solutions I and III) are compatible with those of BESIII within uncertainties. In the fitting procedure of the Belle collaboration the mass and width of $\phi(2170)$ are fixed to the central values obtained by the BESIII collaboration when studying $e^+ e^-\to \phi\eta$ in Ref.~\cite{BESIII:2021bjn}. Based on their results, the Belle collaboration estimated an upper limit for $\mathcal{B}^{\phi(2170)}_{\phi\eta}\Gamma^{\phi(2170)}_{e^+e^-}$ with $90\%$ confident level, being 0.17 eV (for two of the fits) or 18.6 eV (for the remaining two fits). However, the estimated statistical significance for the signal of $\phi(2170)$ in the data of Ref.~\cite{Belle:2022fhh} is low, $1.9\sigma$, and fits to the data without considering $\phi(2170)$ did not seem to produce a difference in the quality of the line shape of the $e^+e^-\to\phi\eta$ cross section. 

In view of the different solutions obtained by the experimental collaborations, a large uncertainty in the determination of the ratio of the decay widths of $\phi(2170)$ to $\phi\eta$, $\Gamma^{\phi(2170)}_{\phi\eta}$, and $\phi\eta^\prime$, $\Gamma^{\phi(2170)}_{\phi\eta^\prime}$ is clearly expected. This could be related to the fact that the cross section for the $e^+ e^-\to\phi\eta$ process in the region of $\simeq 2$ GeV is dominated by the signal of $\phi(1680)$ and its large width ($\simeq 369$ MeV~\cite{BESIII:2021bjn}): as can be seen in the data of Ref.~\cite{BESIII:2021bjn} (which below 2 GeV correspond to that of Ref.~\cite{BaBar:2007ceh}), the signal for $\phi(2170)$ looks like a small bump on top of a background, which is dominated by the tail of the $\phi(1680)$ resonance. This certainly can lead to difficulties in determining the decay width of $\phi(2170)\to \phi \eta$ with precision from fits to the data.

In the last years, several model calculations on the partial decay widths of $\phi(2170)$ have been done by considering different inner structures for the state, producing a broad range of results for the ratio $R_{\eta/\eta^\prime}\equiv \Gamma^{\phi(2170)}_{\phi\eta}/\Gamma^{\phi(2170)}_{\phi\eta^\prime}$~\cite{Ding:2007pc,Li:2020xzs,Barnes:2002mu,Pang:2019ttv,Wang:2021gle,Ding:2006ya,Page:1998gz,Wang:2006ri,Chen:2008ej,Ke:2018evd,Barnes:1995hc,Deng:2010zzd,Chen:2018kuu,Wang:2019nln,Liu:2020lpw}.  The same models predicted ratios of the partial decay widths of $\phi(2170)$ to $\bar K K_R$, where $K_R$ represents a kaonic resonance, like $K(1460)$, $K_1(1270)$, $K_1(1400)$, etc., which are not compatible with the values determined by using the recent results for $\mathcal{B}^{\phi(2170)}_{\bar K K_R}\Gamma^{\phi(2170)}_{e^+e^-}$~\cite{BESIII:2020vtu}. In this way, practically all the models describing $\phi(2170)$ as a $s\bar s$ state, or as a hybrid, or as a tetraquark state have been challenged. It is then important to discuss whether such models provide, or not, a reliable prediction for $R_{\eta/\eta^\prime}$. As we will argue, calculating such ratio might not help in distinguishing between some of the different inner structures proposed for $\phi(2170)$, but it can certainly serve to question some of them. 

Inspired by the growing interest in the experimental determination of the decay widths of $\phi(2170)$ to $\phi\eta$ and $\phi\eta^\prime$, in this work we calculate such widths by considering $\phi(2170)$ as a $\phi K\bar K$ state obtained when $K\bar K$ generates $f_0(980)$~\cite{MartinezTorres:2008gy}. Within such a description, not only the mass and width of $\phi(2170)$ are obtained, with results which are compatible within uncertainties with the experimental ones, but also the cross section data on $e^+e^-\to \phi f_0(980)$ are well reproduced~\cite{Malabarba:2020grf}. Ratios of decay widths of $\phi(2170)$ to $\bar K K_R$ have also been showing to be compatible with those found by using the available data~\cite{Malabarba:2020grf}, including the suppression of the decay mode $\phi(2170)\to\bar K^*(892) K^*(892)$. Such an agreement between theory and experiment is highly non-trivial which implies that our result for $R_{\eta/\eta^\prime}$ must be considered as being reliably determined.

\section{Formalism}\label{for}
\begin{figure}[h!]
\centering
\includegraphics[width=0.5\textwidth]{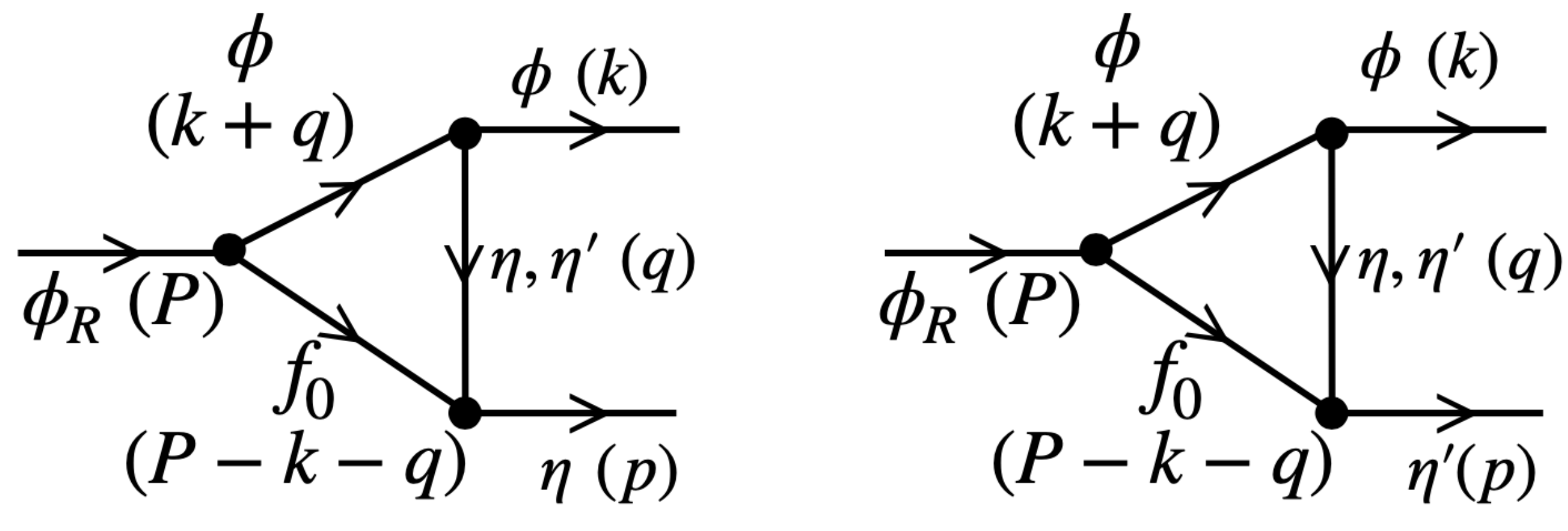}
\caption{Mechanisms for the decay of $\phi_R\equiv\phi(2170)$ as a molecular $\phi f_0(980)$ state to the channels $\phi\eta$ and $\phi\eta^\prime$.}\label{decay}
\end{figure}

Once $\phi(2170)$ is interpreted as a state generated from the interactions in the $\phi K\bar K$ system, with $K\bar K$ forming $f_0(980)$, we need to devise the mechanisms through which a state with such an intrinsic nature could decay to $\phi\eta$ and $\phi\eta^\prime$. As can be seen in Ref.~\cite{ParticleDataGroup:2022pth} (see the review on ``Scalars mesons below 1 GeV''), there seems to accumulate evidence on the molecular nature of $f_0(980)$. We follow Refs.~\cite{Oller:1997ti,Oller:1998hw}, where $f_0(980)$ is generated mainly from the two-body dynamics involved in the $K\bar K$, $\pi\pi$ coupled channel system, in isospin 0, and in the $s$-wave. Although the coupled channel space has been sometimes enlarged by adding the $\eta\eta$ channel due to the proximity of its threshold to the nominal mass of $f_0(980)$~\cite{Oller:1998zr}, channels like $\eta\eta^\prime$ and $\eta^\prime\eta^\prime$ are typically desconsidered for the opposite reason. However, such channels can be incorporated within the formalism of Refs.~\cite{Oller:1997ti,Oller:1998hw} by considering $\eta$ and $\eta^\prime$ as mixtures of a singlet $\eta_1$ and an octet $\eta_8$ of SU(3)~\cite{Herrera-Siklody:1996tqr,Kaiser:2000gs}, with
\begin{align}
|\eta\rangle=\text{cos}\beta|\eta_8\rangle-\text{sin}\beta|\eta_1\rangle,\nonumber\\
|\eta^\prime\rangle=\text{sin}\beta|\eta_8\rangle+\text{cos}\beta|\eta_1\rangle,\label{beta}
\end{align}
where $\beta$ is a mixing angle whose value is in the range $\simeq-15^\circ$ to $-22^\circ$~\cite{Gilman:1987ax,Akhoury:1987ed,Bramon:1997mf,Venugopal:1998fq}. Within this mixing scheme, $f_0(980)$ couples, besides $K\bar K$ and $\pi\pi$, to channels like $\eta\eta$, $\eta\eta^\prime$, $\eta^\prime\eta^\prime$. This allows mechanisms for the decay of $\phi(2170)$  to $\phi\eta$ and $\phi\eta^\prime$ via triangular loops, as shown in Fig.~\ref{decay}.

In this way, knowing the couplings constants of $\phi(2170)$ to $\phi f_0(980)$, and of $f_0(980)$ to $\eta\eta$, $\eta\eta^\prime$, and $\eta^\prime\eta^\prime$, we can determine the contributions from the diagrams in Fig.~\ref{decay} and calculate the decay width of $\phi(2170)$ to $\phi\eta$ and $\phi\eta^\prime$. Calling $t_{\phi_R\to\phi \mathcal{P}}$ to the amplitude for the process $\phi_R\to \phi \mathcal{P}$, where $\phi_R$ is a shorthand notation for $\phi(2170)$ and $\mathcal{P}=\eta,\,\eta^\prime$, the decay width $\Gamma_{\phi_R\to\phi\mathcal{P}}$ is given by
\begin{align}
\Gamma_{\phi_R\to\phi\mathcal{P}}=\int d\Omega \frac{|\vec{p}_{\phi/\mathcal{P}}|}{96\pi^2 m^2_{\phi_R}}\sum\limits_\text{pol}|t_{\phi_R\to\phi \mathcal{P}}|^2.\label{Gamma}
\end{align}
In Eq.~(\ref{Gamma}), $|\vec{p}_{\phi/\mathcal{P}}|$ is the modulus of the linear momentum of the $\phi$ (or $\mathcal{P}$) in the rest frame of the decaying particle, $m_{\phi_R}$ represents the mass of $\phi(2170)$, $\int d\Omega$ represents the integration on the solid angle, and $\sum\limits_\text{pol}$ indicates the sum over the polarizations of the particles in the initial and final states. A factor $1/3$ is included from the average over the polarizations of the initial state.

Using the Feynman rules, the amplitude $t_{\phi_R\to\phi\mathcal{P}}$ can be written as
\begin{align}
&-i t_{\phi_R\to\phi\mathcal{P}}=\sum\limits_{\mathcal{P}^\prime}\int\limits_{-\infty}^{\infty}\frac{d^4q}{(2\pi)^4}(-i t_{\phi_R\to\phi f_0})\nonumber\\
&\quad\times\frac{i}{(P-k-q)^2-m^2_{f_0}+i\epsilon}\nonumber\\
&\quad\times\frac{i}{(k+q)^2-m^2_\phi+i\epsilon}\frac{i}{q^2-m^2_{\mathcal{P}^\prime}+i\epsilon}\nonumber\\
&\quad\times(-i t_{\phi\to \phi \mathcal{P}^\prime})(-i t_{f_0\mathcal{P}^\prime\to\mathcal{P}}),\label{tprocess}
\end{align}
where $\mathcal{P}^\prime=\eta,\,\eta^\prime$, $t_{\phi_R\to \phi f_0}$, $t_{\phi\to \phi \mathcal{P}^\prime}$, $t_{f_0\mathcal{P}^\prime\to\mathcal{P}}$ are the amplitudes describing the transitions $\phi(2170)\to \phi f_0(980)$, $\phi\to \phi\mathcal{P}^\prime$ and $f_0\mathcal{P}^\prime\to\mathcal{P}$, respectively, and $m_\phi$, $m_{f_0}$ and $m_{\mathcal{P}^\prime}$ represent the masses of $\phi$, $f_0(980)$ and $\mathcal{P}^\prime$, respectively. Considering $\phi(2170)$ as a state obtained from the interaction of a $\phi$ with a $f_0(980)$ in the $s$-wave, the amplitude $t_{\phi_R\to\phi f_0}$ can be written as
\begin{align}
t_{\phi_R\to\phi f_0}=g_{\phi_R\to \phi f_0}\epsilon_{\phi_R}(P)\cdot \epsilon_\phi(k+q),
\end{align}
where $g_{\phi_R\to \phi f_0}$ is the coupling constant of $\phi(2170)$ to $\phi f_0(980)$. This coupling constant can be obtained from the model of Ref.~\cite{MartinezTorres:2008gy} by considering that, for energies around the mass of $\phi(2170)$ and invariant masses of the $K\bar K$ system around the mass of $f_0(980)$, the three-body $T$-matrix for $\phi (K\bar K)_0\to \phi (K\bar K)_0$, where the subscript 0 in $K\bar K$ represents isospin, is proportional to the two-body $t$-matrix of $\phi f_0(980)\to \phi f_0(980)$. This proportionality factor can be determined by imposing the unitarity relation for the two-body $t$-matrix, i.e., $\text{Im}[t^{-1}_{\phi f_0\to \phi f_0}]=|\vec{p}^{\,\text{CM}}_{\phi/ f_0}|/(8\pi\sqrt{s})$, with $s$ being the Mandelstam variable of the system and $|\vec{p}^\text{CM}_{\phi/ f_0}|$ the modulus of the linear momentum of $\phi$ (or $f_0$) in the center-of-mass frame~\cite{MartinezTorres:2008gy,Malabarba:2020grf}. The partial decay width of $\phi(2170)\to\phi\mathcal{P}$ depends on $|g_{\phi_R\to \phi f_0}|$, whose value, using the model of Ref.~\cite{MartinezTorres:2008gy}, as given in Ref.~\cite{Malabarba:2020grf} is
\begin{align}
|g_{\phi_R\to \phi f_0}|=3123\pm 561~\text{MeV}.\label{gphiR}
\end{align}
Using this result, a good reproduction of the $e^+ e^-\to \phi f_0(980)$ data was achieved, as shown in Ref.~\cite{Malabarba:2020grf}.

In case of  $t_{f_0\mathcal{P}^\prime\to\mathcal{P}}$, since $f_0(980)$ can be generated from the $s$-wave interaction of two pseudoscalars, we can write the amplitude $t_{f_0\mathcal{P}^\prime\to\mathcal{P}}$ simply as the coupling constant of $f_0(980)$ to the $\mathcal{P}\bar{\mathcal{P}^\prime}$ system, i.e,

\begin{align}
t_{f_0\mathcal{P}^\prime\to\mathcal{P}}=g_{f_0\to \mathcal{P}\bar{\mathcal{P}^\prime}}.\label{tf01}
\end{align}
This coupling constant is obtained from the residue of the corresponding two-body $\mathcal{P}\bar{\mathcal{P}^\prime}$ $t$-matrix in the complex energy plane. The latter $t$-matrix is determined by solving the Bethe-Salpeter equation in its on-shell factorization form~\cite{Oller:1998hw,Liang:2014tia,Lin:2021isc},
\begin{align}
t=(1-VG)^{-1} V.\label{BS}
\end{align}
In Eq.~(\ref{BS}), $V$ and $G$ are matrices (the latter being diagonal) in the coupled channel space and their elements are, respectively, the amplitudes $V_{ij}$ for the process $\mathcal{P}_i\bar{\mathcal{P}^\prime}_i\to\mathcal{P}_j\bar{\mathcal{P}^\prime}_j$ and the loop functions $G_k$ for the two pseudoscalar mesons being propagated in the intermediate channel $k$~\cite{Oller:1998hw}, 
\begin{align}
G_k(\sqrt{s})=\int\limits_0^{Q_\text{max}}dQ\frac{Q^2}{(2\pi)^2}\frac{\omega_{\mathcal{P}_k}+\omega_{\bar{\mathcal{P}^\prime}_k}}{\omega_{\mathcal{P}_k}\omega_{\bar{\mathcal{P}^\prime}_k}[s-(\omega_{\mathcal{P}_k}+\omega_{\bar{\mathcal{P}^\prime}_k})+i\epsilon]},\label{Gk}
\end{align}
with $\omega_{\mathcal{P}_k/\bar{\mathcal{P^\prime}}_k}(Q)=\sqrt{Q^2+m^2_{\mathcal{P}_k/\bar{\mathcal{P^\prime}}_k}}$. This $G_i$ can be regularized through the dimensional regularization scheme or with a cutoff $Q_\text{max}$ for the center of mass momentum $Q=|\vec{Q}|$. Values of $Q_\text{max}\sim 1000$ MeV lead to a good reproduction of the $\pi\pi$ phase-shift in isospin 0, among other observables, up to $\sqrt{s}\simeq 1200$ MeV~\cite{Oller:1997ti, Oller:1998hw}. 

The $V_{ij}$ amplitudes mentioned above are determined from the lowest-order chiral Lagrangian $\mathcal{L}_{\mathbb{P}\mathbb{P}}$, considering the $\eta-\eta^\prime$ mixing~\cite{Herrera-Siklody:1996tqr,Kaiser:2000gs,Liang:2014tia},
\begin{align}
\mathcal{L}_{\mathbb{P}\mathbb{P}}=\frac{1}{12 f^2_\pi}\langle (\partial_\mu \mathbb{P}\mathbb{P}-\mathbb{P}\partial_\mu\mathbb{P})^2+M\mathbb{P}^4\rangle.\label{LPP}
\end{align}
In Eq.~(\ref{LPP}), $f_\pi$ is the weak decay constant of a pion, $f_\pi\simeq 93$ MeV, $\mathbb{P}$ is a matrix containing the fields related to the pseudoscalar mesons,
\begin{widetext}
\begin{align}
\mathbb{P}&=\left(\begin{array}{ccc}A(\beta)\eta+B(\beta)\eta^\prime+\frac{\pi^0}{\sqrt{2}}&\pi^+&K^+\\\pi^-&A(\beta)\eta+B(\beta)\eta^\prime-\frac{\pi^0}{\sqrt{2}}&K^0\\K^-&\bar K^0&C(\beta)\eta+D(\beta)\eta^\prime\end{array}\right),\label{Pmat}
\end{align}
\end{widetext}
where
\begin{align}
A(\beta)&=-\frac{\text{sin}\beta}{\sqrt{3}}+\frac{\text{cos}\beta}{\sqrt{6}},\nonumber\\
B(\beta)&=\frac{\text{sin}\beta}{\sqrt{6}}+\frac{\text{cos}\beta}{\sqrt{3}},\nonumber\\
C(\beta)&=-\frac{\text{sin}\beta}{\sqrt{3}}-\sqrt{\frac{2}{3}}\text{cos}\beta,\nonumber\\
D(\beta)&=-\sqrt{\frac{2}{3}}\text{sin}\beta+\frac{\text{cos}\beta}{\sqrt{3}},
\end{align}
and $M$ is a matrix having as elements
\begin{align}
M&=\left(\begin{array}{ccc}m^2_\pi&0&0\\0&m^2_\pi&0\\0&0&2m^2_K-m^2_\pi\end{array}\right),
\end{align}
where $m_\pi$, $m_K$ represent the masses of the pion and of the kaon, respectively. The convention $-i V=i\mathcal{L}$ has been followed when applying the Feynman rules.

When solving Eq.~(\ref{BS}), we include the pseudoscalar-pseudoscalar channels $K\bar K$, $\pi\pi$, $\eta\eta$, $\eta\eta^\prime$, $\eta^\prime\eta^\prime$ as coupled channels  and consider $\eta-\eta^\prime$ mixing angles $\beta$ between $-15^\circ$ to $-22^\circ$, instead of assuming ideal mixing ($\text{sin}\beta=-1/3$, thus $\beta\simeq -19.47^\circ$)~\cite{Lin:2021isc}.  The elements of $V$ in Eq.~(\ref{BS}) are projected on the $s$-wave, producing a function of the Mandelstam variable $s$ of the system,  the weak decay constants of the pseudoscalars and the masses of the particles involved in the process. When calculating the pseudoscalar-pseudoscalar $t$-matrix, we consider two cases: (I) $V$ determined with different weak decay constants for the pseudoscalars, with $f_\pi=93$ MeV, $f_K=113$ MeV, $f_\eta=f_{\eta^\prime}=111$ MeV, and $G$ obtained with $Q_\text{max}=1020$ MeV~\cite{Oller:1998hw}; (II) $V$ determined with $f_\pi=f_K=f_\eta=f_{\eta^\prime}=93$ MeV and $G$ calculated with $Q_\text{max}=700$ MeV~\cite{Liang:2014tia,Xie:2014tma}. As shown in the appendix, these two cases enfold the experimental results, including the error bars, on the $s$-wave $\pi\pi$ phase shift in isospin 0, where $f_0(980)$ is observed. Note that for the case (I), the factor $f_\pi^2$ appearing in $V_{ij}$  needs to be replaced by $\sqrt{f_{\mathcal{P}_i} f_{\bar{\mathcal{P}^\prime}_i}f_{\mathcal{P}_j} f_{\bar{\mathcal{P}^\prime}_j}}$~\cite{Oller:1998hw}. In Table~\ref{f0coup} we give the pole positions and coupling constants of $f_0(980)$ to the $\eta\eta$, $\eta\eta^\prime$, $\eta^\prime\eta^\prime$ channels for three different mixing angles and refer the reader to the appendix~\ref{f0ap} for more details. 
\begin{table}[h!]
\caption{Pole position and coupling constant $g_i$ of $f_0(980)$ to the channel $i$ for three mixing angles, including ideal mixing, $\beta\simeq -19.47^\circ$. The notation $M-i\Gamma/2$, with $M$ being the mass of the resonance and $\Gamma$ its width, is used to denote the pole obtained. The first (second) row in each entry represents the result obtained for case (I) [(II)].\\}\label{f0coup}
\begin{tabular}{cccc}
$\beta$ (Degrees)& $-15$& $-19.47$ & $-22$\\\\
\multirow{2}{*}{$\text{Pole}$ (MeV)}&$979.7 -i 11.0$&$ 984.10-i10.3$&$986.3-i9.7$\\
&$971.6-i9.8$&$975.3-i9.6$&$977.4-i9.4$ \\\\
\multirow{2}{*}{$g_{\eta\eta}$ (MeV)}&$-3529-i 261$ &$-3169-i 346$ &$-2966-i 405$\\
&$-3972-i142$&$-3615-i178$&$-3413-i201$ \\\\
\multirow{2}{*}{$g_{\eta\eta^\prime}$ (MeV)}&$1534+i 101$ & $ 1490+i 145$ &$1443+i 177$\\
&$1677+i45$&$1628+i61$&$1588+i72$\\\\
\multirow{2}{*}{$g_{\eta^\prime\eta^\prime}$ (MeV)}&$-2057-i 187$ & $-2127-i 261$ & $-2152-i 318$\\
&$-2270-i118$&$-2369-i149$&$-2412-i170$
\end{tabular}
\end{table}

Next, we need to determine $t_{\phi\to\phi\mathcal{P}^\prime}$. This amplitude, which involves two vector and a pseudoscalar mesons, can be obtained by using the Lagrangian~\cite{Fujiwara:1984mp,Bando:1985rf,Bramon:1994pq}
\begin{align}
\mathcal{L}_{VVP}=\frac{g_{VVP}}{\sqrt{2}}\epsilon^{\mu\nu\alpha\beta}\langle \partial_\mu V_\nu\partial_\alpha V_\beta \mathbb{P}\rangle,\label{LVVP}
\end{align}
where $\mathbb{P}$ is given by Eq.~(\ref{Pmat}), $g_{VVP}=3m^2_V/(16\pi^2 f^3_\pi)$, with $m_V\simeq m_\rho$, $f_\pi\simeq 93$ MeV, and 
\begin{align}
V_\mu&=\left(\begin{array}{ccc}\frac{\omega+\rho^0}{\sqrt{2}}&\rho^+&K^{*+}\\\rho^-&\frac{\omega-\rho^0}{\sqrt{2}}&K^{*0}\\K^{*-}&\bar K^{*0}&\phi\end{array}\right).
\end{align}
Considering Eq.~(\ref{LVVP}) and the Feynman rules, we get the following $t_{\phi\to\phi\mathcal{P}^\prime}$ amplitude,
\begin{align}
t_{\phi\to\phi\mathcal{P}^\prime}=2 g_{\phi\to\phi\mathcal{P}^\prime}\epsilon^{\mu\nu\alpha\beta}q_\mu k_\alpha\epsilon_{\phi\nu}(k+q)\epsilon_{\phi\beta}(k),\label{tVVP}
\end{align}
with $g_{\phi\to\phi\mathcal{P}^\prime}=-\frac{g_{VVP}}{\sqrt{2}} \mathcal{C}_{\mathcal{P}^\prime}$, where
\begin{align}
 \mathcal{C}_{\mathcal{P}^\prime}=\left\{\begin{array}{c}-\frac{1}{\sqrt{3}}\text{sin}\beta-\sqrt{\frac{2}{3}}\text{cos}\beta,~\text{for}~\mathcal{P}^\prime=\eta,\\-\sqrt{\frac{2}{3}}\text{sin}\beta+\frac{1}{\sqrt{3}}\text{cos}\beta,~\text{for}~\mathcal{P}^\prime=\eta^\prime.\end{array}\right.
\end{align}
To obtain Eq.~(\ref{tVVP}) we have made used of the antisymmetric properties of the Levi-Civita tensor $\epsilon^{\mu\nu\alpha\beta}$.

Using the above amplitudes, we can write Eq.~(\ref{tprocess}) as
\begin{align}
&it_{\phi_R\to\phi\mathcal{P}}=-\sum\limits_{\mathcal{P}^\prime} g_{\phi_R\to\phi f_0}g_{f_0\to\mathcal{P}\bar{\mathcal{P}^\prime}}g_{\phi\to\phi\mathcal{P}^\prime}\epsilon^{\mu^\prime}_{\phi_R}(P)\nonumber\\
&\quad\times\int\limits_{-\infty}^{\infty}\frac{d^4q}{(2\pi)^4}\epsilon^{\mu\nu\alpha\beta}q_\mu k_\alpha\epsilon_{\phi\beta}(k)\Bigg[-g_{\mu^\prime\nu}\nonumber\\
&\quad+\frac{(k+q)_{\mu^\prime}(k+q)_\nu}{m^2_\phi}\Bigg]\frac{1}{(P-k-q)^2-m^2_{f_0}+i\epsilon}\nonumber\\
&\quad\times\frac{1}{(k+q)^2-m^2_\phi+i\epsilon}\frac{1}{q^2-m^2_{\mathcal{P}^\prime}+i\epsilon},\label{tprocesssum}
\end{align}
where a sum over the polarizations of the internal vector meson has been carried out.  Using the antisymmetric properties of the Levi-Civita tensor, Eq.~(\ref{tprocesssum}) can be written as
\begin{align}
it_{\phi_R\to\phi\mathcal{P}}&=\sum\limits_{\mathcal{P}^\prime}2g_{\phi_R\to\phi f_0}g_{f_0\to\mathcal{P}\bar{\mathcal{P}^\prime}}g_{\phi\to\phi\mathcal{P}^\prime}\nonumber\\
&\quad\times\epsilon^{\mu\nu\alpha\beta}\epsilon_{\phi_R\nu}(P)k_\alpha\epsilon_{\phi\beta}(k)I_\mu,\label{tImu}
\end{align}
where $I_\mu$ represents the tensor integral
\begin{align}
I_\mu&=\int\limits_{-\infty}^{\infty}\frac{d^4q}{(2\pi)^4}\frac{q_\mu}{[(P-k-q)^2-m^2_{f_0}+i\epsilon]}\nonumber\\
&\quad\times\frac{1}{[(k+q)^2-m^2_\phi+i\epsilon][q^2-m^2_{\mathcal{P}^\prime}+i\epsilon]}.\label{Imudef}
\end{align}

To calculate $I_\mu$ we proceed as follows: from Eq.~(\ref{Imudef}), after integrating on $d^4 q$,  the integral $I_\mu$ can depend on two quadrimomenta, $k_\mu$ and $P_\mu$. Using the Lorentz covariance, we can write then $I_\mu$ as
\begin{align}
I_\mu=a_{\mathcal{P}^\prime} k_\mu+b_{\mathcal{P}^\prime} P_\mu,\label{Imu}
\end{align}
where $a_{\mathcal{P}^\prime}$ and $b_{\mathcal{P}^\prime}$ are unknown coefficients to be determined. These coefficients can be expressed in terms of two scalar integrals and they depend on the mass of $\mathcal{P}^\prime$ (besides other quantities). Indeed, by multiplying Eq.~(\ref{Imu}) by $k^\mu$ and $P^\mu$, respectively, we obtain a system of two coupled equations which permits to express the coefficients $a_{\mathcal{P}^\prime}$ and $b_{\mathcal{P}^\prime}$ as
\begin{align}
a_{\mathcal{P}^\prime}&=\frac{P^2(k\cdot I)-(k\cdot P)(P\cdot I)}{k^2 P^2-(k\cdot P)^2},\nonumber\\
b_{\mathcal{P}^\prime}&=-\frac{(k\cdot P)(k\cdot I)-k^2(P\cdot I)}{k^2 P^2-(k\cdot P)^2},
\end{align}
with 
\begin{align}
k\cdot I&=\int\limits_{-\infty}^{\infty}\frac{d^4q}{(2\pi)^4}\frac{k\cdot q}{[(P-k-q)^2-m^2_{f_0}+i\epsilon]}\nonumber\\
&\quad\times\frac{1}{[(k+q)^2-m^2_\phi+i\epsilon][q^2-m^2_{\mathcal{P}^\prime}+i\epsilon]},\nonumber\\
P\cdot I&=\int\limits_{-\infty}^{\infty}\frac{d^4q}{(2\pi)^4}\frac{P\cdot q}{[(P-k-q)^2-m^2_{f_0}+i\epsilon]}\nonumber\\
&\quad\times\frac{1}{[(k+q)^2-m^2_\phi+i\epsilon][q^2-m^2_{\mathcal{P}^\prime}+i\epsilon]}.
\end{align}
Next, we must mention that we work in the rest frame of the decaying particle, thus, $P^\mu=(P^0,\vec{0})$, with $P^0=m_{\phi_R}$. In this way, we can write the preceding integrals as
\begin{align}
k\cdot I&=\int\limits_{-\infty}^{\infty}\frac{d^3q}{(2\pi)^3}\int\limits_{-\infty}^\infty\frac{dq^0}{(2\pi)}\frac{k^0q^0-\vec{k}\cdot\vec{q}}{[(P-k-q)^2-m^2_{f_0}+i\epsilon]}\nonumber\\
&\quad\times\frac{1}{[(k+q)^2-m^2_\phi+i\epsilon][q^2-m^2_{\mathcal{P}^\prime}+i\epsilon]}\nonumber\\
&\equiv \int\limits_{-\infty}^{\infty}\frac{d^3q}{(2\pi)^3}[k^0\mathcal{I}_1(m_{f_0},m_\phi,m_{\mathcal{P}^\prime})\nonumber\\
&\quad-\vec{k}\cdot\vec{q}\,\mathcal{I}_0(m_{f_0},m_\phi,m_{\mathcal{P}^\prime})],\nonumber\\
P\cdot I&=P^0\int\limits_{-\infty}^{\infty}\frac{d^3q}{(2\pi)^3}\int\limits_{-\infty}^\infty\frac{dq^0}{(2\pi)} \frac{q^0}{[(P-k-q)^2-m^2_{f_0}+i\epsilon]}\nonumber\\
&\quad\times\frac{1}{[(k+q)^2-m^2_\phi+i\epsilon][q^2-m^2_{\mathcal{P}^\prime}+i\epsilon]}\nonumber\\
&\equiv P^0\int\limits_{-\infty}^{\infty}\frac{d^3q}{(2\pi)^3}\mathcal{I}_1(m_{f_0},m_\phi,m_{\mathcal{P}^\prime}),\label{QI}
\end{align}
where we have introduced
\begin{align}
&\mathcal{I}_n(m_1,m_2,m_3)\equiv\int\limits_{-\infty}^\infty\frac{dq^0}{(2\pi)}\frac{(q^0)^n}{[(P-k-q)^2-m^2_1+i\epsilon]}\nonumber\\
&\quad\times\frac{1}{[(k+q)^2-m^2_2+i\epsilon][q^2-m^2_3,+i\epsilon]}\label{Icaln}
\end{align}
with $n=0,1$. The integral in Eq.~(\ref{Icaln}) can be obtained by separating explicitly the $q^0$ dependence in the denominator and by using Cauchy's theorem, finding
\begin{align}
\mathcal{I}_n(m_1,m_2,m_3)=-i\frac{N_n(m_1,m_2,m_3)}{D(m_1,m_2,m_3)},\label{lcalnrat}
\end{align}
where 
\begin{align}
&D(m_1,m_2,m_3)=2 E_1 E_2 E_3(P^0+E_1+E_2)(k^0+E_2+E_3)\nonumber\\
&\quad \times(P^0-E_1-E_2+i\epsilon)(P^0-k^0-E_1-E_3+i\epsilon)\nonumber\\
&\quad\times(-P^0+k^0-E_1-E_3+i\epsilon)(k^0-E_2-E_3+i\epsilon),
\end{align}
\begin{align}
&N_0(m_1,m_2,m_3)=(E_1+E_2)\Big[(E_1+E_3)(E_2+E_3)\nonumber\\
&\quad\times(E_1+E_2+E_3)-E_3 (k^0)^2\Big]\nonumber\\
&\quad-(P^0)^2 E_1(E_2+E_3)+2 E_1 E_3 k^0 P^0,
\end{align}
\begin{align}
&N_1(m_1,m_2,m_3)=-E_3\Big[k^0(E_1+E_2)\{E_1(E_1+E_2\nonumber\\
&\quad+2 E_3)+(E_2+E_3)^2-(k^0)^2\}+P^0\{(k^0)^2\nonumber\\
&\quad\times (2E_1+E_2)-E_2(E_2+E_3)(2E_1\nonumber\\
&\quad+E_2+E_3)\}-(P^0)^2 k^0 E_1\Big],
\end{align}
with $E_1=\sqrt{(\vec{k}+\vec{q}\,)^2+m^2_1}$, $E_2=\sqrt{(\vec{k}+\vec{q}\,)^2+m^2_2}$, $E_3=\sqrt{\vec{q}^{\,2}+m^2_3}$. The integral in $d^3 q$ in Eq.~(\ref{QI}) can be obtained as
\begin{align}
\int\limits_{-\infty}^\infty \frac{d^3q}{(2\pi)^3}(\cdots)&\to\int\limits_0^\infty \frac{d|\vec{q}\,| |\vec{q}\,|^2}{(2\pi)^2}\int\limits_{-1}^{1}\text{dcos}\theta F(\Lambda,|\vec{k}+\vec{q}|)\nonumber\\
&\quad\times F(\bar \Lambda,|\vec{q}^{\,\text{CM}}_{\mathcal{P}/\bar{\mathcal{P}^\prime}}|)(\cdots),\label{d3q}
\end{align}
where we choose $\vec{k}=|\vec{k}|\hat {z}$, and $\vec{q}=|\vec{q}\,|\text{sin}\theta(\text{cos}\phi\hat{i}+\text{sin}\phi\hat{j})+|\vec{q}\,|\text{cos}\theta\hat{k}$, such that $\vec{k}\cdot\vec{q}=|\vec{k}| |\vec{q}\,|\text{cos}\theta$ and the integral in $d\phi$ is trivial. In Eq.~(\ref{d3q}), $F$ is a form-factor introduced for the vertices $\phi_R\to\phi f_0$ and $f_0\mathcal{P}^\prime\to\mathcal{P}$ and $\Lambda$, $\bar \Lambda$ are cutoffs for the center-of-mass momentum of the particles forming these states. It is important to mention here that although the integrals are convergent, form factors at the vertices of $\phi_R\to\phi f_0$ and $f_0\to\mathcal{P}\bar{\mathcal{P}}^\prime$ are introduced to take into account the finite size of $\phi(2170)$ and $f_0(980)$. In case of the vertex $\phi(2170)\to \phi f_0$ a value of $\Lambda\simeq m_{\phi}+m_{f_0}\simeq 2000$ MeV is considered, while for the vertex $f_0(980)\mathcal{P}^\prime\to\mathcal{P}$ the value of $\bar\Lambda$ is related to the cutoff used to regularize $G$ in Eq.~(\ref{BS}) when generating $f_0(980)$ from the interaction of two pseudoscalar mesons in $s$-wave, i.e., $\bar \Lambda\simeq 1000$ MeV. Note that the modulus of the linear momentum of $\mathcal{P}$ (or $\bar{\mathcal{P}^\prime}$) in the rest frame of $f_0(980)$, $|\vec{q}^{\,\text{CM}}_{\mathcal{P}/\mathcal{P}^\prime}|$, is related to $\vec{k}$ and $\vec{q}$ through a boost from a frame where $f_0(980)$ has linear momentum $-(\vec{k}+\vec{q})$ to the one in which $f_0(980)$ is at rest. Typical expressions for the form factors in Eq.~(\ref{d3q}) are Lorentz~\cite{Gamermann:2009uq},
\begin{align}
F(\Lambda,|\vec{Q}|)=\frac{\Lambda^2}{\Lambda^2+|\vec{Q}|^2},
\end{align}
or Gaussian functions,
\begin{align}
F(\Lambda,|\vec{Q}|)=e^{-\frac{|\vec{Q}|^2}{2\Lambda^2}}.
\end{align}
For a better comparison of the results obtained by using different form factors, the latter are normalized in such a way that~\cite{Gamermann:2009uq} 
\begin{align}
&\int_0^{Q_\text{max}}d|\vec{Q}|\Theta(Q_\text{max}-|\vec{Q}|)\Theta(Q_\text{max}-|\vec{Q}|)\nonumber\\
&\quad=\int_0^{\Lambda_L}d|\vec{Q}|F^2_L(|\vec{Q}|,\Lambda_L)\nonumber\\
&\quad=\int_0^{\Lambda_G}d|\vec{Q}|F^2_G(|\vec{Q}|,\Lambda_G)\label{relat}
\end{align}
where $Q_\text{max}$ coincides with the cutoff of Eq.~(\ref{Gk}). The subscripts $L$, $G$ in Eq.~(\ref{relat}) represent the Lorentz and Gaussian functional form of the form factors.
 
 Once we have established how to calculate the integrals appearing in the coefficients $a_{\mathcal{P}^\prime}$ and $b_{\mathcal{P}^\prime}$, by using Eqs.~(\ref{tImu}) and (\ref{Imu}),
 we can write
 \begin{align}
 it_{\phi_R\to\phi\mathcal{P}}&=\sum\limits_{\mathcal{P}^\prime}2 g_{\phi_R\to\phi f_0}g_{f_0\to\mathcal{P}\bar{\mathcal{P}^\prime}} g_{\phi\to\phi\mathcal{P}^\prime}\nonumber\\
 &\quad\times\epsilon^{\mu\nu\alpha\beta}\epsilon_{\phi_R\nu}(P)k_\alpha\epsilon_{\phi\beta}(k)P_\mu b_{\mathcal{P}^\prime}.\label{tfinal}
 \end{align}
Note that in Eq.~(\ref{tfinal}) the term of $I_\mu$ related to $a_{\mathcal{P}^\prime}$ does not contribute as a consequence of the antisymmetric properties of the Levi-Civita tensor. 

To determine the decay width $\Gamma_{\phi_R\to\phi\mathcal{P}}$, we need to calculate $\sum\limits_\text{pol}|t_{\phi_R\to\phi\mathcal{P}}|^2$. Using Eq.~(\ref{tfinal}), we have
\begin{align}
&\sum\limits_\text{pol}|t_{\phi_R\to\phi\mathcal{P}}|^2=\sum\limits_{\mathcal{P}^\prime_a,\mathcal{P}^\prime_b}4|g_{\phi_R\to\phi f_0}|^2g_{f_0\to \mathcal{P}\bar{\mathcal{P}}^\prime_a}\nonumber\\
&\quad\times g^*_{f_0\to\mathcal{P}\bar{\mathcal{P}}^\prime_b}g_{\phi\to\phi\mathcal{P}^\prime_a}g^*_{\phi\to\phi\mathcal{P}^\prime_b}\epsilon^{\mu\nu\alpha\beta}\epsilon_{\mu^\prime\nu^\prime\alpha^\prime\beta^\prime}\nonumber\\
&\quad\times\left(-g^{\nu^\prime}_\nu+\frac{P^{\nu^\prime}P_\nu}{m^2_\phi}\right)k_\alpha\left(-g^{\beta^\prime}_\beta+\frac{k^{\beta^\prime}k_\beta}{m^2_\phi}\right)\nonumber\\
&\quad\times P_\mu b_{\mathcal{P}^\prime_a}b^*_{\mathcal{P}^\prime_b}k^{\alpha^\prime}P^{\mu^\prime},
\end{align}
where $\mathcal{P}^\prime_a$, $\mathcal{P}^\prime_b=\eta$, $\eta^\prime$. Once again, using the antisymmetric properties of the Levi-Civita tensor, we obtain
\begin{align}
&\sum\limits_\text{pol}|t_{\phi_R\to\phi\mathcal{P}}|^2=\sum\limits_{\mathcal{P}^\prime_a,\mathcal{P}^\prime_b}4|g_{\phi_R\to\phi f_0}|^2g_{f_0\to\mathcal{P}\bar{\mathcal{P}}^\prime_a}g^*_{f_0\to\mathcal{P}\bar{\mathcal{P}}^\prime_b}\nonumber\\
&\quad\times g_{\phi\to\phi\mathcal{P}^\prime_a}g^*_{\phi\to\phi\mathcal{P}^\prime_b}\epsilon^{\mu\nu\alpha\beta}\epsilon_{\mu^\prime\nu\alpha^\prime\beta}k_\alpha P_\mu k^{\alpha^\prime}\nonumber\\
&\quad\times P^{\mu^\prime}b_{\mathcal{P}^\prime_a}b^*_{\mathcal{P}^\prime_b}.\label{tpol}
\end{align}
Next, considering the property of the Levi-Civita tensor
\begin{align}
\epsilon^{\mu\nu\alpha\beta}\epsilon_{\mu^\prime\nu^\prime\alpha^\prime\beta^\prime}=-\left|\begin{array}{cccc}g^\mu_{\mu^\prime}&g^\mu_{\nu^\prime}&g^\mu_{\alpha^\prime}&g^\mu_{\beta^\prime}\\g^\nu_{\mu^\prime}&g^\nu_{\nu^\prime}&g^\nu_{\alpha^\prime}&g^\nu_{\beta^\prime}\\g^\alpha_{\mu^\prime}&g^\alpha_{\nu^\prime}&g^\alpha_{\alpha^\prime}&g^\alpha_{\beta^\prime}\\g^\beta_{\mu^\prime}&g^\beta_{\nu^\prime}&g^\beta_{\alpha^\prime}&g^\beta_{\beta^\prime}\end{array}\right|,\label{epep}
\end{align}
the product of the tensors appearing in Eq.~(\ref{tpol}) can be written as
\begin{align}
\epsilon^{\mu\nu\alpha\beta}\epsilon_{\mu^\prime\nu\alpha^\prime\beta}=2(g^\mu_{\alpha^\prime}g^\alpha_{\mu^\prime}-g^\mu_{\mu^\prime}g^\alpha_{\alpha^\prime}).\label{epep}
\end{align}
Using Eq.~(\ref{epep}) we obtain the following expression for $\sum\limits_\text{pol}|t_{\phi_R\to\phi\mathcal{P}}|^2$:
\begin{align}
&\sum\limits_\text{pol}|t_{\phi_R\to\phi\mathcal{P}}|^2=\sum\limits_{\mathcal{P}^\prime_a,\mathcal{P}^\prime_b}8|g_{\phi_R\to\phi f_0}|^2g_{f_0\to\mathcal{P}\bar{\mathcal{P}}^\prime_a}g^*_{f_0\to\mathcal{P}\bar{\mathcal{P}}^\prime_b}\nonumber\\
&\quad\times g_{\phi\to\phi\mathcal{P}^\prime_a}g^*_{\phi\to\phi\mathcal{P}^\prime_b}[(k\cdot P)^2-k^2 P^2]b_{\mathcal{P}^\prime_a}b^*_{\mathcal{P}^\prime_b}.
\end{align}
Expanding explicitly the sum in the preceding equation, we obtain
\begin{align}
&\sum\limits_\text{pol}|t_{\phi_R\to\phi\mathcal{P}}|^2=8|g_{\phi_R\to\phi f_0}|^2\Bigg[|g_{f_0\to\mathcal{P}\eta}|^2|g_{\phi\to\phi\eta}|^2|b_\eta|^2\nonumber\\
&\quad+2\text{Re}\{g_{f_0\to\mathcal{P}\eta }g^*_{f_0\to\mathcal{P}\eta^\prime}g_{\phi\to\phi\eta}g^*_{\phi\to\phi\eta^\prime}b_\eta b^*_{\eta^\prime}\}\nonumber\\
&\quad+|g_{f_0\to\mathcal{P}\eta^\prime}|^2|g_{\phi\to\phi\eta^\prime}|^2|b_{\eta^\prime}|^2\Bigg][(k\cdot P)^2-k^2 P^2].\label{tpolsqf}
\end{align}
This is the expression which is used for calculating the decay width of $\phi(2170)\to\phi \eta$, $\phi\eta^\prime$ through Eq.~(\ref{Gamma}). Since in the rest frame of the decaying particle $k\cdot P=k^0 P^0$, with $P^0=m_{\phi_R}$ and
\begin{align}
k^0=\frac{m^2_{\phi_R}+m^2_\phi-m^2_{\mathcal{P}}}{2m_{\phi_R}},
\end{align}
the expression in Eq.~(\ref{tpolsqf}) does not depend on the solid angle present in Eq.~(\ref{Gamma}). In this way, we can carry the integral on $d\Omega$ explicitly and write
\begin{align}
\Gamma_{\phi_R\to\phi\mathcal{P}}=\frac{|\vec{p}_{\phi/\mathcal{P}}|}{24\pi m^2_{\phi_R}}\sum\limits_\text{pol}|t_{\phi_R\to\phi \mathcal{P}}|^2,\label{wfor}
\end{align}
with $\sum\limits_\text{pol}|t_{\phi_R\to\phi \mathcal{P}}|^2$ given by Eq.~(\ref{tpolsqf}).

\section{Results and discussions}
Let us now discuss the results of our calculations. In table~\ref{widths} we show the decay widths of $\phi(2170)$ to $\phi\eta$ and $\phi\eta^\prime$ obtained by using different $\eta-\eta^\prime$ mixing angles and form factors, as mentioned in Sec.~\ref{for}. In particular, we consider Lorentz and Gaussian form factors. As can be seen, the values found with different mixing angles as well as different form factors are compatible within error bars. The central value and the uncertainty in the results shown in Table~\ref{widths} represent, respectively, the mean and standard deviation obtained for the widths when generating random numbers for the coupling of $\phi(2170)$ to $\phi f_0$ within the interval established by Eq.~(\ref{gphiR}). 
\begin{table}[h!]
\caption{Decay widths (in MeV) of $\phi(2170)$ to $\phi\eta$ and $\phi\eta^\prime$ for different $\eta-\eta^\prime$ mixing angles, $\beta$, and different form factors. The labels L and G indicate the consideration of a Lorentz (L) or Gaussian (G) form factors, while the numbers I and II refer to the model used to calculate the $\mathcal{P}\bar{\mathcal{P}}^\prime$ $t$-matrix.\\}\label{widths}
\begin{tabular}{ccccc}
$\beta$ (Degree)&&$-15$&$-19.47$&$-22$\\
\\
\multirow{4}{*}{$\Gamma^{\phi(2170)}_{\phi\eta}$}&LI&$4.30\pm0.93$&$3.27\pm 0.71$ &$2.76\pm 0.60$ \\ 
&GI&$5.14\pm 1.11$&$3.91\pm 0.85$&$3.29\pm 0.71$\\
&LII&$3.38\pm0.73$&$2.59\pm0.56$&$2.20\pm0.48$\\
&GII&$4.17\pm0.91$&$3.20\pm0.69$&$2.71\pm0.59$
\\\\
\multirow{4}{*}{$\Gamma^{\phi(2170)}_{\phi\eta^\prime}$}&LI&$0.84\pm 0.18$&$0.83\pm 0.18$&$0.81\pm 0.18$\\
&GI&$0.94\pm 0.20$&$0.93\pm0.20$&$0.91\pm0.20$\\
&LII&$0.80\pm0.18$&$0.80\pm0.17$&$0.79\pm0.17$\\
&GII&$0.95\pm0.21$&$0.94\pm0.20$&$0.93\pm0.20$ 
\end{tabular}
\end{table}

The results of Table~\ref{widths} show several interesting facts. For instance, the decay width obtained for the process $\phi(2170)\to\phi\eta$ is relatively small when compared to the total width of $\phi(2170)$ ($\simeq 50-100$ MeV), in spite of the large phase space available ($m_\phi+m_\eta\simeq1547$ MeV). At the same time, the decay mode of the state to the $\phi\eta^\prime$ channel, for which there is less phase space available for decaying ($m_\phi+m_{\eta^\prime}\simeq 1978$ MeV), although having a smaller decay width than that for $\phi\eta$, is not very much suppressed. We also find that the decay widths reduce when increasing the modulus value of the $\eta-\eta^\prime$ mixing angle. These properties are a direct consequence of the nature of $\phi(2170)$ as a $\phi f_0(980)$ molecular state, of $f_0(980)$ as a state generated from the interaction of two pseudoscalars and of the vector-vector-pseudoscalar vertices, which altogether establish the decay mechanisms of $\phi(2170)$ to $\phi\eta$ and $\phi\eta^\prime$ shown in Fig.~\ref{decay}.

Using the results of Table~\ref{widths}, we list in Table~\ref{ratio} the values found for the ratio $R_{\eta/\eta^\prime}=\Gamma^{\phi(2170)}_{\phi\eta}/\Gamma^{\phi(2170)}_{\phi\eta^\prime}$ considering different mixing angles and form factors. A comment is here in order before continuing further with the discussions. Within the theoretical approach developed, it is not expected that $R_{\eta/\eta^\prime}$ depends on the coupling constant $g_{\phi_R\to\phi f_0}$. This is so because the latter coupling, as can be seen in Fig.~\ref{decay}, is involved in the primary vertex of the decay mechanisms represented in Fig.~\ref{decay} and, thus, it should cancel when calculating $R_{\eta/\eta^\prime}$. However, for a better comparison of our result for $R_{\eta/\eta^\prime}$ with the value obtained by using experimental data as $\mathcal{B}^{\phi(2170)}_{\phi\eta}\Gamma^{\phi(2170)}_{e^+e^-}/\mathcal{B}^{\phi(2170)}_{\phi\eta^\prime}\Gamma^{\phi(2170)}_{e^+e^-}$, we consider the results given in Table~\ref{widths} for $\Gamma^{\phi(2170)}_{\phi\eta}$ and $\Gamma^{\phi(2170)}_{\phi\eta^\prime}$ as being independent from each other. This is inline with the fact that the values of $\mathcal{B}^{\phi(2170)}_{\phi\eta}\Gamma^{\phi(2170)}_{e^+e^-}$ and $\mathcal{B}^{\phi(2170)}_{\phi\eta^\prime}\Gamma^{\phi(2170)}_{e^+e^-}$ were extracted from fits to the data on the cross section of the processes $e^+ e^-\to \phi \eta$~\cite{BESIII:2021bjn,Belle:2022fhh} and of $e^+ e^-\to \phi\eta^\prime$~\cite{BESIII:2020gnc}, respectively, finding different nominal values for the mass and width of $\phi(2170)$. In view of this, and since $g_{\phi_R\to\phi f_0}$ depends on the mass and width of $\phi(2170)$, we find it reasonable to implement the propagation of errors when calculating $R_{\eta/\eta^\prime}$ from the results of Table~\ref{widths}.

\begin{table}[h!]
\caption{Values for the ratio $R_{\eta/\eta^\prime}$ between $\Gamma^{\phi(2170)}_{\phi\eta}$ and $\Gamma^{\phi(2170)}_{\phi\eta^\prime}$ for different $\eta-\eta^\prime$ mixing angles, $\beta$, and form factors. The meaning of the labels is the same as in Table~\ref{widths}.\\}\label{ratio}
\begin{tabular}{ccccc}
$\beta$ (Degree)&&$-15$&$-19.47$&$-22$\\\\
\multirow{4}{*}{$R_{\eta/\eta^\prime}$}&LI&$5.12\pm 1.57$&$3.93\pm 1.21$ &$3.39\pm 1.04$ \\ 
&GI&$5.47\pm 1.68$&$4.21\pm 1.29$&$3.63\pm 1.11$\\
&LII&$4.21\pm1.29$&$3.25\pm1.00$&$2.80\pm0.86$\\
&GII&$4.41\pm1.35$&$3.40\pm1.04$&$2.93\pm0.90$
\end{tabular}
\end{table}

Calculating the average and standard deviation of the values of $R_{\eta/\eta^\prime}$ shown in Table~\ref{ratio}, we get
\begin{align}
R_{\eta/\eta^\prime}=3.9\pm1.3.\label{our}
\end{align}

We can compare this result with the ratio $R^\text{exp}_{\eta/\eta^\prime}\equiv \mathcal{B}^{\phi(2170)}_{\phi\eta}\Gamma^{\phi(2170)}_{e^+e^-}/\mathcal{B}^{\phi(2170)}_{\phi\eta^\prime}\Gamma^{\phi(2170)}_{e^+e^-}$ obtained by using the values $\mathcal{B}^{\phi(2170)}_{\phi\mathcal{P}}\Gamma^{\phi(2170)}_{e^+e^-}$ found in Refs.~\cite{BESIII:2021bjn,BESIII:2020gnc}:
\begin{align}
R^\text{exp}_{\eta/\eta^\prime}=\left\{\begin{array}{l}0.034^{+0.018}_{-0.011}~\text{solution I},\\ 1.42^{+0.58}_{-0.48}~\text{solution II,}\end{array}\right.\label{Rbes}
\end{align}
and with the results obtained for $R^\text{exp}_{\eta/\eta^\prime}$ by using for $\mathcal{B}^{\phi(2170)}_{\phi\eta}\Gamma^{\phi(2170)}_{e^+e^-}$ the value found in Ref.~\cite{Belle:2022fhh}, which gives
\begin{align}
R^\text{exp}_{\eta/\eta^\prime}=\left\{\begin{array}{l}0.013\pm0.007~\text{solution I},\\ 0.009\pm 0.003~\text{solution II},\\2.4\pm0.4~\text{solutions III, IV}.\end{array}\right.\label{RBB}
\end{align}
As can be seen, we get, in general, values of $R_{\eta/\eta^\prime}$ which are higher than those extracted from the experimental data. However, our lower limit of $2.6$ for $R_{\eta/\eta^\prime}$ in Eq.~(\ref{our}) is close to the upper limit of $2$ for solution II in Eq.~(\ref{Rbes}) as well as to the upper limit of 2.8 for solutions III and IV of Eq.~(\ref{RBB}). Such an agreement alone is not trivial to find, but together with the results obtained in Ref.~\cite{Malabarba:2020grf} is remarkable.

At this point, it is important to emphasize that the properties of $\phi(2170)$ obtained from the independent fits to the data on Ref.~\cite{BESIII:2021bjn} and to that of Ref.~\cite{BESIII:2020gnc} are not very compatible. In these fits, it is considered that that the line shapes of the cross sections $e^+e^-\to\phi\eta$, $e^+e^-\to\phi\eta^\prime$ are represented by a coherent sum of phase space and a Breit-Wigner, the latter depending on the mass and width of $\phi(2170)$ as well as of the product $\Gamma^{\phi(2170)}_{e^+e^-}\mathcal{B}^{\phi(2170)}_{\phi\mathcal{P}}$, with $\mathcal{P}$ being $\eta$ or $\eta^\prime$. In Ref.~\cite{BESIII:2021bjn}, the mass and width extracted for $\phi(2170)$ from fits to data on the cross section of $e^+e^-\to\phi\eta$ are of $2163.5\pm 6.2\pm 3$ MeV and $31.1^{+21.1}_{-11.6}\pm 1.1$ MeV, respectively. But in Ref.~\cite{BESIII:2020gnc}, the corresponding mass and width obtained from fits to the data on the cross section of $e^+ e^-\to\phi\eta^\prime$ are, respectively, $2177.5\pm 4.8\pm 19.5$ MeV and $149.0\pm 15.6\pm 8.9$. While the masses obtained in Refs.~\cite{BESIII:2021bjn} and \cite{BESIII:2020gnc} are compatible within uncertainties, the widths differ significantly, even when considering the corresponding uncertainties.  The reason for such large deviations is still unclear, but it could affect the determination of $R_{\eta/\eta^\prime}$ with the present data. Further, in Ref.~\cite{Belle:2022fhh}, where the Belle collaboration studied the process $e^+e^-\to\phi\eta$, the mass and width for $\phi(2170)$ in the fits is fixed to the values found in Ref.~\cite{BESIII:2021bjn}. 

In view of such discrepancies on the properties of $\phi(2170)$, dividing the value of $\mathcal{B}^{\phi(2170)}_{\phi\eta}\Gamma^{\phi(2170)}_{e^+e^-}$ obtained in Ref.~\cite{BESIII:2021bjn} with the one for 
$\mathcal{B}^{\phi(2170)}_{\phi\eta^\prime}\Gamma^{\phi(2170)}_{e^+e^-}$ found in Ref.~\cite{BESIII:2020gnc} may not be the best way of predicting the ratio $\Gamma^{\phi(2170)}_{\phi\eta}/\Gamma^{\phi(2170)}_{\phi\eta^\prime}$. In spite of that, it is worth stressing once more that our lowest value for $R_{\eta/\eta^\prime}$ is close to the highest values found with some of the solutions of Refs.~\cite{BESIII:2020gnc,BESIII:2021bjn,Belle:2022fhh}. Determination of  experimental data with higher precision is necessary to deduce a precise value of $R_{\eta/\eta^\prime}$.

\subsection{Comparison with other models}

It is also important to compare our result for $R_{\eta/\eta^\prime}$ with the results found in other theoretical models trying to understand the properties and nature of $\phi(2170)$. 

\subsubsection{$\phi(2170)$ as a $s\bar s$ state}
In Ref.~\cite{Barnes:2002mu}, using the ${}^3P_0$ decay model with simple harmonic oscillator $q\bar q$ wave functions, a $3^3 S_1$ $s\bar s$ vector state, also denoted as $\phi(3S)$, with an estimated mass of $2050$ MeV and a width of $\simeq 380$ MeV was predicted (here the spectroscopic notation $n^{2s+1}L_J$ is used to denote the $n$th state with spin $s$, orbital angular momentum $L$ and total angular momentum $J$ for a quark-antiquark system, with $L=0,1,2,\dots$ being represented by the letters $S$, $P$, $D$, etc.). Such a state has a decay width to $\phi\eta$ of 21 MeV, while its decay width to $\phi\eta^\prime$ is 11 MeV, leading to a ratio of the two to be $\simeq 1.91$ MeV. Although the value obtained for $R_{\eta/\eta^\prime}$ within the model of Ref.~\cite{Barnes:2002mu} is of similar order to ours, the individual widths found are much larger. Interestingly, the value of $R_{\eta/\eta^\prime}$ obtained within the model of Ref.~\cite{Barnes:2002mu} is close to the upper (lower) limit found for solution II (solutions III and IV) in Eq.~(\ref{Rbes}) [Eq.~(\ref{RBB})]. However, in view of the discrepancy of the mass and width obtained for the $3^3 S_1$ $s\bar s$ state in Ref.~\cite{Barnes:2002mu} with those of $\phi(2170)$, associating the ratio $R_{\eta/\eta^\prime}$ found for such state with that for $\phi(2170)$ is not very meaningful.  

In Ref.~\cite{Ding:2007pc}, the possibility of $\phi(2170)$ being a $2^3 D_1$ $s\bar s$ meson [also denoted as $\phi (2D)$] and its decay modes were studied by considering the ${}^3 P_0$ and the flux tube models. Although the decay widths of $\phi(2170)$ to different channels, like $\bar KK(1460)$, $\bar K^*(892) K^*(892)$, depend strongly on the parameters of the model, it is concluded that such state can either decay strongly to $\bar K K(1460)$, $\bar K K^*(1410)$, $\bar K K_1(1270)$, $\bar K^*(892) K^*(892)$ and $\bar KK$ or to $\bar K K_1(1400)$, $\eta h_1(1380)$, $\bar K^*(892) K^*(892)$, $\bar K K_1(1270)$. Both situations are ruled out by the experimental findings of Ref.~\cite{BESIII:2020vtu}, where the decay mode of $\phi(2170)$ to $K^{*+}(892)K^{*-}(892)$ is suppressed and the corresponding partial decay widths to $K^- K^+_1(1400)$ and $K^- K^+_1(1270)$ are comparable. These facts, however, can be understood within the a $\phi f_0$ description for $\phi(2170)$, as shown in Ref.~\cite{Malabarba:2020grf}. Besides, with the chosen model parameters, the decay width to $\phi\eta$ is found to be zero, while for $\phi\eta^\prime$ is $\simeq 3$ MeV.

In Ref.~\cite{Pang:2019ttv}, by using a modified version of the Godfrey-Isgur model~\cite{Godfrey:1985xj}, a  $3^3 S_1$ $s\bar s$ state, with a mass of 2149 MeV, and a $2^3 D_1$ $s\bar s$ state, with a mass of 2276 MeV, are obtained. In this way, $\phi(2170)$ lies between these two states. The decay widths found for the $\phi(3S)$ state to $\phi\eta$ and $\phi\eta^\prime$ by considering a mass of 2188 MeV for the state are, respectively, 6.66 and 0.0862 MeV, thus $R_{\eta/\eta^\prime}\simeq 77$. In case of the $\phi(2D)$ state, a decay width of 0.879 MeV (0.0887 MeV) is obtained for the $\phi\eta$ ($\phi\eta^\prime$) channel, having then $R_{\eta/\eta^\prime}\simeq 9.9$.  The values found for $R_{\eta/\eta^\prime}$ for the $\phi(3S)$ and $\phi(2D)$ states are not compatible with those of Eqs.~(\ref{Rbes}), (\ref{RBB}). As already mentioned by the authors of Ref.~\cite{Pang:2019ttv}, the width found for the $\phi(3S)$ state is of 217 MeV, which is not compatible with that of $\phi(2170)$, while for the $\phi(2D)$ state, the width obtained is of 186 MeV, with a significant fraction (18$\%$) of this value corresponding to its decay to $K^*(892)\bar K^*(892)$. This latter finding is not supported by the results obtained by the BESIII collaboration in Ref.~\cite{BESIII:2020vtu}, where the decay of $\phi(2170)\to K^*(892)\bar K^*(892)$ is found to be suppressed.  

In Ref.~\cite{Li:2020xzs}, the $s\bar s$ mass spectrum is determined within a nonrelativistic linear potential quark model and strong decay widths are evaluated using the ${}^3 P_0$ model. A $3^3 S_1$ $s\bar s$ vector state with a mass of 2198 MeV and a width of $\sim 240-270$ MeV is obtained, with $K^*(892)\bar K^*(892)$ and $K\bar K^*(1410)$ being the main decay modes. Such properties are not compatible with those observed in Ref.~\cite{BESIII:2020vtu}. In any case, considering a mass of 2175 MeV for the $3^3 S_1$ $s\bar s$ state, its decay width to $\phi\eta$ is $\simeq 8.9$ MeV, while to $\phi\eta^\prime$ the value obtained is 0.36 MeV. In this way, $R_{\eta/\eta^\prime}\simeq 25$, which is not compatible with the results inferred from Refs.~\cite{BESIII:2021bjn,BESIII:2020gnc,Belle:2022fhh}. Assigning lower masses to $\phi(3S)$ (in the range $2079-2135$ MeV) leads to $R_{\eta/\eta^\prime}\geq 100$. 

In Ref.~\cite{Li:2020xzs}, a $\phi(2D)$ with mass in the interval 2050-2200 MeV is also considered. In case of the mass being $\simeq$2175 MeV, the width of the $\phi(2D)$ state is too broad ($\sim 300$ MeV) to associate such a state with $\phi(2170)$. Besides this, the main decay modes are $K\bar K^*(1410)$ and $K^*(892)\bar K^*(892)$, findings which are not compatible with those of Ref.~\cite{BESIII:2020vtu}. The value obtained for $R_{\eta/\eta^\prime}$ is the same as in case of the $\phi(3S)$ state. If, however, the mass of the $\phi(2D)$ state is considered to be in the range 2079-2135 MeV, a smaller width is found ($\simeq 175-225$ MeV), but the partial decay width to $K^*(892)\bar K^*(892)$ is still large, comparable with that to the $K \bar K_1(1400)$ channel. Also, $\phi(2170)\to K^*(892)\bar K^*(892)$ is still one of the main decay modes, which is not compatible with the properties obtained in Ref.~\cite{BESIII:2020vtu}. For completeness, it is worth mentioning that in the latter case the decay width to $\phi\eta$ is 0.34 (0.64) MeV and to $\phi\eta^\prime$ is of 0.3 (0.23) MeV for a mass of the state of 2079 (2135) MeV, thus, $R_{\eta/\eta^\prime}\simeq 1.13$ (2.8). Since the masses and widths of the $\phi(3S)$ and $\phi(2D)$ states are in the same range, and both states have similar strong decay properties, the authors of Ref.~\cite{Li:2020xzs} argue about the possibility that the experimental signal observed for $\phi(2170)$ could be a superposition of the $\phi(3S)$ and $\phi(2D)$ states. In spite of this, the authors of Ref.~\cite{Li:2020xzs} assert that such mixing should not reduce the decay width of the state to $K^*(892)\bar K^*(892)$, and  the strong suppression of the decay mode of $\phi(2170)$ to $K^*(892)\bar K^*(892)$ observed in Ref.~\cite{BESIII:2020vtu} would remain a mystery within the formalism of Ref.~\cite{Li:2020xzs}. 

In Ref.~\cite{Wang:2021gle} the mass spectrum and decay behavior of excited states of $\rho$, $\omega$ and $\phi$ mesons above 2 GeV are determined by using an unquenched potential model. A $\phi(3S)$ state with a mass of 2103 MeV and a width of 156 MeV is found, as well as a $\phi(2D)$ state with 2236 MeV of mass and 265 MeV of width. None of these results match the nominal mass and width of $\phi(2170)$. However, by using some effective Lagrangians and by fixing the coupling constants of the $\rho$, $\omega$ and $\phi$ states obtained to $\bar K K_R$ from experimental data, the authors of Ref.~\cite{Wang:2021gle} can describe the experimental data on the cross sections of $e^+e^-\to K^+K^-$, $\bar K K^*_2(1430)+\text{cc.}$, $K^-K^+_1(1270)$, $K^{*+}(892)K^{*-}(892)$ obtained by the BaBar and BESIII collaborations. Note, however, that to do that, the mass and width of the $\phi(3S)$ and $\phi(2D)$ states are considered as free parameters in the fitting procedure. The values of mass and width found for these $\phi$ states from the fits are $m_{\phi(3S)}=2183\pm 1$ MeV, $\Gamma_{\phi(3S)}=185\pm 4$ MeV, $m_{\phi(2D)}=2290\pm 3$ MeV, $\Gamma_{\phi(2D)}=312\pm 6$ MeV. As can be seen, the masses, as well as the widths, of the $\phi(3S)$ and $\phi(2D)$ states obtained from the fit differ significantly from their theoretical results. The authors of Ref.~\cite{Wang:2021gle} interpret the disagreement between the nominal mass and width of $\phi(2170)$ extracted from the data and those of the $\phi(3S)$ and $\phi(2D)$ states obtained within their model in terms of the interference between the latter states. However, in view of the discrepancy between the theoretical results obtained for the masses and widths of these states and those needed to explain the data considered, it is not clear if such interference is responsible for the signal observed for $\phi(2170)$.

\subsubsection{$\phi(2170)$ as a $s\bar sg$ hybrid}
The state $\phi(2170)$ has also been interpreted as a $s\bar s g$ hybrid, with a mass between 2100-2200 MeV and a width of $\simeq 120-170$ MeV~\cite{Barnes:1995hc,Page:1998gz,Ding:2006ya}. In Ref.~\cite{Page:1998gz}, the ratio $R_{\eta/\eta^\prime}$ for a $1^{--}$ $s\bar s$ hybrid with mass $\simeq 2000-2200$ MeV changes significantly with the parameters of the model, ranging from $9.5-200$. These values are not compatible with those found in Eqs.~(\ref{Rbes}), (\ref{RBB}). In case of the model of Ref.~\cite{Ding:2006ya}, the decay widths of the state to $\phi\eta$ and $\phi\eta^\prime$ are 1.2 and 0.4 MeV, respectively. In this way, a value of $3$ is obtained for $R_{\eta/\eta^\prime}$, inline with our result. However, in the models of Refs.~\cite{Page:1998gz,Ding:2006ya}, the decay mode of such a hybrid state to $\bar K K(1460)$ is highly suppressed. This fact is in contradiction with the experimental result of Ref.~\cite{BESIII:2020vtu} as well as with the one obtained by considering $\phi(2170)$ as a $\phi f_0$ state~\cite{Malabarba:2020grf}. We can conclude then that simply measuring the ratio $R_{\eta/\eta^\prime}$ might not be enough to distinguish between a $s\bar s g$ and a $\phi f_0$ inner structures for $\phi(2170)$.

In Ref.~\cite{Ma:2020bex}, the formation of $s\bar s g$ hybrids has been investigated from lattice QCD in the quenched approximation by considering spatially extended $s\bar s$ and $s\bar s g$ operators and no conclusive information could be obtained for the state $\phi(2170)$. The authors of Ref.~\cite{Ma:2020bex}, however, argue that, independently of whether $\phi(2170)$ is a $s\bar s$ meson or a $s\bar sg$ hybrid, the ratio $R_{\eta/\eta^\prime}$ is expected to be the same in both cases since the same decay dynamics of the state to $\phi\eta$ and $\phi\eta^\prime$ can be presumed. In this way, the ratio $R_{\eta/\eta^\prime}$ is determined by the flavor-octet-singlet mixing angle, producing the physical $\eta$ and $\eta^\prime$ particles, and the corresponding phase space kinematical factors. By considering mixing angles between $-10^\circ$ and $-20^\circ$, values of $R_{\eta/\eta^\prime}\simeq 0.14-0.58$ are obtained. These values are compatible with the result $0.23\pm0.10\pm 0.18$, which is obtained by dividing the value of $\mathcal{B}^{\phi(2170)}_{\phi\eta}\Gamma^{\phi(2170)}_{e^+e^-}$ available from the BaBar collaboration~\cite{BaBar:2006gsq} by the value for $\mathcal{B}^{\phi(2170)}_{\phi\eta^\prime}\Gamma^{\phi(2170)}_{e^+e^-}$ of the BESIII collaboration~\cite{BESIII:2020gnc}. The value $R_{\eta/\eta^\prime}\simeq 0.14-0.58$, however, is not compatible with the ratios  shown in Eqs.~(\ref{Rbes}) and (\ref{RBB}), determined using the data from the BESIII/Belle collaboration. Nevertheless, the authors of Ref.~\cite{Ma:2020bex} conclude that the ratio $R_{\eta/\eta^\prime}$ may not be the most useful quantity to distinguish a $s\bar s$ nature for $\phi(2170)$ from a hybrid $s\bar sg$. 

\subsubsection{$\phi(2170)$ as a tetraquark state}
In Ref.~\cite{Wang:2006ri} a QCD sum rule analysis is made to study the possible tetraquark $ss\bar s\bar s$ nature of $\phi(2170)$. By considering the standard sum rule convergence criteria, and a rather small Borel window, a mass of $2460\pm 160$ MeV was found, which is larger than the nominal value of $\phi(2170)$. The author of Ref.~\cite{Wang:2006ri} considers then a more phenomenological approach and a mass of $2210\pm 90$ MeV is obtained. Neither the full width nor the partial decay widths of $\phi(2170)$ were obtained in Ref.~\cite{Wang:2006ri}. Such information is essential to know if the state obtained can be, or not, related to $\phi(2170)$.

In Ref.~\cite{Chen:2008ej} a QCD sum rule analysis of the nature of $\phi(2170)$ by considering diquark-antidiquark $(ss)(\bar s\bar s)$ and meson-meson $(\bar s s)(\bar s s)$ currents is done. Two independent interpolating currents where found and a mass of $2300\pm 400$ MeV is obtained for $\phi(2170)$. Note that  the central value for the mass of $\phi(2170)$ is about 130 MeV higher than the nominal one, although as a consequence of the large uncertainty obtained in the mass sum rule, their result could be compatible. Neither the full width nor the partial decay widths of $\phi(2170)$ were determined in Ref.~\cite{Chen:2008ej}. 

By considering a flux tube model, the possible $ss\bar s\bar s$ tetraquark nature of $\phi(2170)$ was investigated in Ref.~\cite{Deng:2010zzd} by solving the four-body Schr\"odinger equation. Based on the symmetry of the wave function of the tetraquark system under the exchange of two identical quarks, two possible rearrangements for the quarks were obtained in Ref.~\cite{Deng:2010zzd}. In one of this configuration, a mass of 2290 MeV was determined for $\phi(2170)$, while for the other configuration a mass of $2188$ MeV was found. Considering this latter result, the authors of Ref.~\cite{Deng:2010zzd} associated a tetraquark nature with $\phi(2170)$. However, apart from the mass, no other observables which could be compared with the experimental data to support such an assignment were determined in Ref.~\cite{Deng:2010zzd}. 

In Ref.~\cite{Chen:2018kuu}, the calculation done in Ref.~\cite{Chen:2008ej} was improved by considering two different mixtures between two interpolating currents. For one of the mixtures considered, a mass of $2410\pm 250$ MeV is obtained, while for the other mixture, a mass of $2340\pm 170$ MeV is found. The authors associate the latter result with $\phi(2170)$ and use the former one to suggest the existence of a partner for $\phi(2170)$. Note, however, that it is basically the lower limit of the mass obtained for $\phi(2170)$ in Ref.~\cite{Chen:2018kuu} the value which is compatible with the experimental result and the central value found in Ref.~\cite{Chen:2018kuu} is still far from the nominal mass of $\phi(2170)$. Also, no other observables, like full width, partial decay widths, etc., were determined in Ref.~\cite{Chen:2018kuu}. Thus it is not clear if the tetraquark state obtained in Ref.~\cite{Chen:2018kuu} with mass $2340\pm 170$ MeV can be related to $\phi(2170)$.

By assuming a tetraquark nature for $\phi(2170)$, the authors of Ref.~\cite{Ke:2018evd} studied the decay modes of $\phi(2170)$ in terms of the so-called fall-apart mechanism~\cite{Jaffe:1976ih}. By considering a $s s\bar s\bar s$ nature for $\phi(2170)$, the authors arrived to the conclusion that the decay widths of $\phi(2170)$ to $\bar K K_1(1270)$ and $\bar K K_1(1400)$ should be similar, while the decay to $\bar K^*(892) K^*(892)$ should be suppressed. These facts seem to be compatible with those observed by the BESIII collaboration in Ref.~\cite{BESIII:2020vtu}. The ratios of the widths of $\phi(2170)$ to $\phi\eta$, $\phi\eta^\prime$, $\phi f_0(980)$ are determined to be $\Gamma^{\phi(2170)}_{\phi\eta}: \Gamma^{\phi(2170)}_{\phi\eta^\prime}:\Gamma^{\phi(2170)}_{\phi f_0(980)}\sim 0.015:0.025:1$, thus a value of $\simeq 0.6$ is obtained for $R_{\eta/\eta^\prime}$. Such a value is close to the lower limit of $0.94$ of the solution II of Eq.~(\ref{Rbes}). Similarly, the authors of Ref.~\cite{Ke:2018evd} also estimated the ratios between some partial decay widths for $\phi(2170)$ by assuming a $q\bar q s\bar s$ nature for $\phi(2170)$, with $q=u,d$, and by considering $f_0(980)$ to be a tetraquark state. 

Continuing with the discussion on the studies investigating the tetraquark nature of $\phi(2170)$, a QCD sum rule analysis performed in Ref.~\cite{Wang:2019nln} for a vector $ss\bar s\bar s$ current produces a mass of $3080\pm 110$ MeV. This mass lies well above the nominal mass of $\phi(2170)$, disfavoring such internal structure for the state. 

In Ref.~\cite{Liu:2020lpw}, the possible $ss\bar s\bar s$ nature of $\phi(2170)$ was investigated within a framework based on a nonrelativistic potential quark model by considering that the four quarks rearrange themselves in a diquark-antidiquark configuration. Masses in the range of $\simeq 2440-2990$ MeV are obtained for $J^{PC}=1^{--}$ tetraquark states. These masses are much higher than the nominal value for $\phi(2170)$ and the authors of Ref.~\cite{Liu:2020lpw} reached the conclusion that $\phi(2170)$ might not be a good candidate for a $ss\bar s\bar s$ state.

\subsubsection{$\phi(2170)$ as a $\Lambda \bar\Lambda$ state and estimation of the $\phi\eta$, $\phi\eta^\prime$ decay widths}
The $\phi(2170)$ state has also been interpreted as a $\Lambda\bar\Lambda$ bound state~\cite{Zhao:2013ffn,Dong:2017rmg}. In Ref.~\cite{Zhao:2013ffn}, using a one-boson exchange model, where $\phi$, $\omega$, $\sigma$, $\eta$ and $\eta^\prime$ exchange was considered, the authors studied the $\Lambda\bar\Lambda$ system by solving the Schr\"odinger equation. The coupling constants characterizing the strength of the meson-$\Lambda$-$\Lambda$ vertex were determined by using the Gell-Mann-Okubo mass formula and SU(3) symmetry, which relates the former couplings to meson-nucleon-nucleon couplings (the latter being determined from the Bonn model~\cite{Machleidt:1987hj}). Two states where found in Ref.~\cite{Zhao:2013ffn}: a ${}^3S_1$ state, with a binding energy of $\simeq  50-83$ MeV, and a ${}^1S_0$ state, whose binding energy is in the range $\simeq 8-13$ MeV. The former was associated with $\phi(2170)$ while the latter was related to $\eta(2225)$. In Ref.~\cite{Dong:2017rmg}, considering the model of Ref.~\cite{Zhao:2013ffn}, the compositeness condition~\cite{Weinberg:1962hj} was used to determine the coupling of $\phi(2170)$ and $\eta(2225)$ to $\Lambda\bar\Lambda$ and several partial decay widths to meson-meson final states were evaluated. In particular, the authors found that $\phi(2170)$ would decay dominantly to $K\bar K$, with a decay width of $\simeq 74-88$ MeV. 
\begin{figure}
\includegraphics[width=0.3\textwidth]{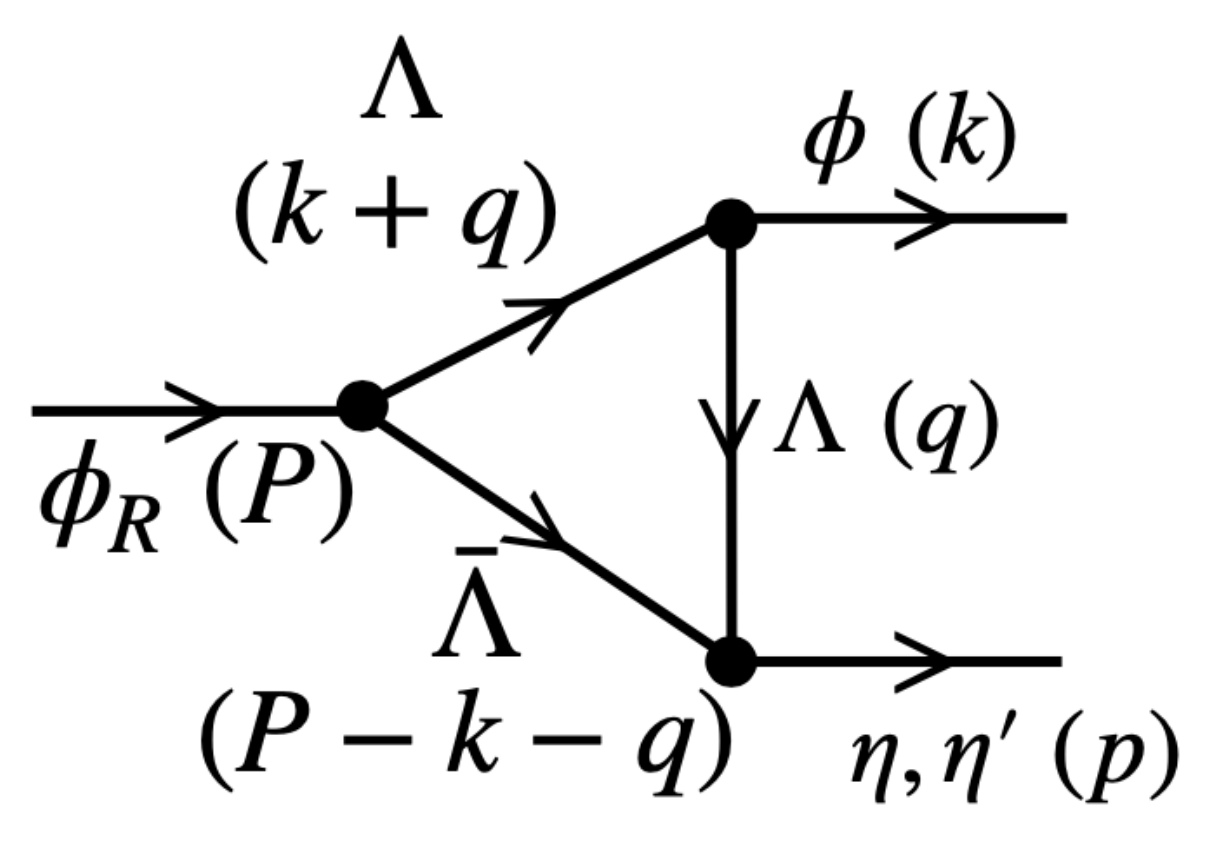}
\caption{$\phi(2170)$ decaying to $\phi\eta$, $\phi\eta^\prime$ within the $\Lambda\bar\Lambda$ description.}\label{phiLambdaLambdabar}
\end{figure}

Since the decay mechanism of $\phi(2170)$ to $\phi\eta$ and $\phi\eta^\prime$ as a $\Lambda\bar\Lambda$ state  (see Fig.~\ref{phiLambdaLambdabar}) proceeds through a triangular loop, as in case of the $\phi f_0(980)$ description, it would be interesting\footnote{We thank the referee for such a suggestion.} to estimate the corresponding decay widths within the model of Ref.~\cite{Dong:2017rmg}. In this way we can compare the ratio $R_{\eta/\eta^\prime}$ with the result extracted from the experimental data, as well as with the value obtained within our model. 

Considering the effective Lagrangians of Ref.~\cite{Dong:2017rmg} to describe the contributions of the vertices involved in Fig.~\ref{phiLambdaLambdabar}, we find the following amplitude for $\phi(2170)\to\phi\mathcal{P}$, with $\mathcal{P}=\eta$, $\eta^\prime$,
\begin{align}
&-i t_{\phi_R\to\phi\mathcal{P}}=i g_{\phi_R\to\Lambda\bar\Lambda}g_{\phi\Lambda\Lambda}g_{\mathcal{P}\Lambda\Lambda}\epsilon^\mu_{\phi_R}(P)\epsilon^\nu_\phi(k)\nonumber\\
&\quad\times\int\limits_{-\infty}^{\infty}\frac{d^4q}{(2\pi)^4}\text{Tr}[(\slashed{k}+\slashed{q}+m_\Lambda)\gamma_\mu(\slashed{P}-\slashed{k}-\slashed{q}-m_\Lambda)\nonumber\\
&\quad\times \gamma_5(\slashed{q}+m_\Lambda)\gamma_\nu]\frac{1}{(P-k-q)^2-m^2_\Lambda+i\epsilon}\nonumber\\
&\quad\times\frac{1}{(k+q)^2-m^2_\Lambda+i\epsilon}\frac{1}{q^2-m^2_\Lambda+i\epsilon},\label{tLL}
\end{align}
where $m_\Lambda$ is the mass of the $\Lambda$ ($\bar\Lambda)$ particle and $g_{\mathcal{P}\Lambda\Lambda}$, $g_{\phi\Lambda\Lambda}$ and $g_{\phi_R\to\Lambda\bar\Lambda}$ are the coupling constants of the vertices involved. Using the properties of the $\gamma$-matrices, we can calculate the trace present in Eq.~(\ref{tLL}) to obtain
\begin{align}
&-it_{\phi_R\to\phi\mathcal{P}}=4m_\Lambda g_{\phi_R\to\Lambda\bar\Lambda}g_{\phi\Lambda\Lambda}g_{\mathcal{P}\Lambda\Lambda}\epsilon_{\mu\nu\alpha\beta}k^\alpha P^\beta \epsilon^\mu_{\phi_R}(P)\nonumber\\
&\quad\times\epsilon^\nu_\phi(k)\int\limits_{-\infty}^\infty\frac{d^4q}{(2\pi)^4}\frac{1}{(P-k-q)^2-m^2_\Lambda+i\epsilon}\nonumber\\
&\quad\times\frac{1}{(k+q)^2-m^2_\Lambda+i\epsilon}\frac{1}{q^2-m^2_\Lambda+i\epsilon}.\label{tLL2}
\end{align}
Using Eqs.~(\ref{Icaln}) and (\ref{lcalnrat}), we can write Eq.~(\ref{tLL2}) as
\begin{align}
-it_{\phi_R\to\phi\mathcal{P}}&=4m_\Lambda g_{\phi_R\to\Lambda\bar\Lambda}g_{\phi\Lambda\Lambda}g_{\mathcal{P}\Lambda\Lambda}\nonumber\\
&\quad\times\epsilon_{\mu\nu\alpha\beta}k^\alpha P^\beta \epsilon^\mu_{\phi_R}(P)\epsilon^\nu_\phi(k)I^{\Lambda}_{\mathcal{P}},
\end{align}
where
\begin{align}
I^{\Lambda}_{\mathcal{P}}\equiv\int\limits_{-\infty}^{\infty} \frac{d^3q}{(2\pi)^3}\mathcal{I}_0(m_\Lambda,m_\Lambda,m_\Lambda).\label{IPLambda}
\end{align}
The integral in Eq.~(\ref{IPLambda}) can be calculated numerically, as explained earlier. In this way, summing over the polarizations of the vector mesons and using Eq.~(\ref{epep}), we find
\begin{align}
\sum\limits_{\text{pol}}|t_{\phi_R\to\phi\mathcal{P}}|^2&=32 m^2_\Lambda|g_{\phi_R\to\Lambda\bar\Lambda}|^2|g_{\phi\Lambda\Lambda}|^2|g_{\mathcal{P}\Lambda\Lambda}|^2\nonumber\\
&\quad\times|I^{\Lambda}_{\mathcal{P}}|^2[(k\cdot P)^2-k^2 P^2].\label{resLL}
\end{align}
Using Eqs.~(\ref{resLL}) and (\ref{wfor}) we can determine the decay width of $\phi(2170)$ to $\phi\eta$ and $\phi\eta^\prime$ by considering $\phi(2170)$ as a $\Lambda\bar\Lambda$ state. The coupling constants appearing in Eq.~(\ref{resLL}) can be found in Refs.~\cite{Zhao:2013ffn,Dong:2017rmg}. The values obtained for the decay widths of $\phi(2170)\to\phi\eta, \phi\eta^\prime$ are respectively $\simeq 17$ and $15.5$ MeV, producing a ratio $R_{\eta/\eta^\prime}\simeq 1.1$. As can be seen from Table~\ref{widths}, the widths determined within the $\Lambda\bar\Lambda$ description are much larger than those found within the $\phi f_0(980)$ model for $\phi(2170)$.  Also, the value determined for the ratio $R_{\eta/\eta^\prime}$ within the former description is smaller (see Table~\ref{ratio}), and it is inside the interval produced by the error bar associated with $R^\text{exp}_{\eta/\eta^\prime}$ (solution II) of Eq.~(\ref{Rbes}). 

Note, however, that $R_{\eta/\eta^\prime}$ depends on the ratio $|g_{\eta\Lambda\Lambda}|/|g_{\eta^\prime\Lambda\Lambda}|$, which at the same time, it is a function of the $\eta-\eta^\prime$ mixing angle~\cite{Zhao:2013ffn}. It can be easily obtained that the values used for the $\mathcal{P}\Lambda\Lambda$ coupling constants in Refs.~\cite{Zhao:2013ffn,Dong:2017rmg} correspond to a $\eta-\eta^\prime$ mixing angle of $\simeq -11^\circ$. For a better determination of $R_{\eta/\eta^\prime}$, $\eta-\eta^\prime$ mixing angles between $\simeq-15^\circ$ to $-22^\circ$ should be considered, although it is not clear to us if such mixing angles are compatible with the Gell-Mann-Okubo mass formula used to derive the coupling constants in Ref.~\cite{Zhao:2013ffn} (see Ref.~\cite{Burakovsky:1997sd}). If we blindly apply the formulas for the $\mathcal{P}\Lambda\Lambda$ couplings obtained in Ref.~\cite{Zhao:2013ffn} for mixing angles in the range $\simeq-15^\circ$ to $-22^\circ$, the decay width of $\phi(2170)\to\phi\eta$ is found to be in the interval $\simeq 10-21$ MeV, while for the decay width of $\phi(2170)\to\phi\eta^\prime$ we obtain $18-33$ MeV, with a ratio $R_{\eta/\eta^\prime}\simeq 0.43-0.86$. This latter value would be, in any case, close to the lower limit $R^\text{exp}_{\eta/\eta^\prime}=0.94$ of Eq.~(\ref{Rbes}) (solution II). 

Before continuing with further discussions, some remarks are here in order. The authors of Ref.~\cite{Dong:2017rmg}, which we follow, consider an effective Lagrangian of the type $g_{\mathcal{P}\Lambda\Lambda}\bar \Psi_\Lambda \gamma_5\Psi_\Lambda\phi_\mathcal{P}$ to describe the $\mathcal{P}\Lambda\Lambda$ vertex, with $\Psi_\Lambda$ and $\phi_\mathcal{P}$ being the corresponding fields. Note, however, that such a vertex structure coincides with that determined from chiral symmetry~\cite{Klingl:1997kf}, i.e., $f_{\mathcal{P}\Lambda\Lambda}\bar \Psi_\Lambda\gamma_\mu\gamma_5\Psi_\Lambda\partial^\mu\phi_\mathcal{P}$, for on-shell baryons, with the coupling $f_{\mathcal{P}\Lambda\Lambda}$ being related to the previous one through $g_{\mathcal{P}\Lambda\Lambda}=2m_\Lambda f_{\mathcal{P}\Lambda\Lambda}$. Obtaining $f_{\mathcal{P}\Lambda\Lambda}$ by considering the $\eta-\eta^\prime$ mixing of Eq.~(\ref{Pmat}) as done in Ref.~\cite{Nagahiro:2011fi}, produces 
\begin{align}
g_{\eta\Lambda\Lambda}&=(2m_\Lambda)\sqrt{\frac{2}{3}}\frac{D}{f_\pi}\Big(\text{sin}\beta+\frac{\text{cos}\beta}{\sqrt{2}}\Big),\nonumber\\
g_{\eta^\prime\Lambda\Lambda}&=(2m_\Lambda)\sqrt{\frac{2}{3}}\frac{D}{f_\pi}\Big(\frac{\text{sin}\beta}{\sqrt{2}}-\text{cos}\beta\Big),\label{gour}
\end{align}
with $D\simeq 0.75-0.8$~\cite{Borasoy:1998pe}. In this way,
\begin{align}
\frac{g_{\eta\Lambda\Lambda}}{g_{\eta^\prime\Lambda\Lambda}}=-\frac{\text{cos}\beta+\sqrt{2}\text{sin}\beta}{\sqrt{2}\text{cos}\beta-\text{sin}\beta},\label{ratour}
\end{align}
instead of $-(\text{cos}\beta+\text{sin}\beta)/(\text{cos}\beta-\text{sin}\beta)$~\cite{Zhao:2013ffn}. The sources of such differences are unknown to us. Using Eq.~(\ref{ratour}) and $\eta-\eta^\prime$ mixing angles in the interval $\simeq-15^\circ$ to $-22^\circ$ we obtain $R_{\eta/\eta^\prime}\simeq 0.14-0.34$, which is not compatible with the values of Eq.~(\ref{Rbes}). 

It should also be mentioned that since the $\Lambda$'s in the decay mechanism depicted in Fig.~\ref{phiLambdaLambdabar} are not on-shell, the use of a $\mathcal{P}\Lambda\Lambda$ vertex structure different to that considered in Ref.~\cite{Dong:2017rmg} could produce a different off-shell behavior for the decay process. This could change the results obtained for the decay widths and the $R_{\eta/\eta^\prime}$ ratio. Such a study is beyond the scope of the estimation presented in this work for $R_{\eta/\eta^\prime}$ by using the model of Refs.~\cite{Zhao:2013ffn,Dong:2017rmg}.

The reader should also be aware that the possible $\Lambda\bar\Lambda$ nature of $\phi(2170)$ is not free of controversy. In Ref.~\cite{Zhu:2019ibc} the $\Lambda\bar\Lambda$ interaction was studied by using the same one-boson-exchange model of Ref.~\cite{Zhao:2013ffn} and by solving a 3-dimensional reduction of the 4-dimensional Bethe-Salpeter equation. As in case of Ref.~\cite{Zhao:2013ffn}, $J^P=1^-$, $0^-$ states are observed, but the mass obtained for the $1^-$ state lies close to the $\Lambda\bar\Lambda$ threshold. The authors of Ref.~\cite{Zhu:2019ibc} related then the $1^-$ state found to $X(2239)$, which was observed by the BESIII collaboration in the cross-sectional data of $e^+e^-\to K^+K^-$~\cite{BESIII:2018ldc}, and not to $\phi(2170)$. Since the authors of Ref.~\cite{Zhu:2019ibc} consider the same input Lagrangians as those of Ref.~\cite{Zhao:2013ffn}, the former authors attributed the discrepancies between the two models to different treatments, such as different method to solve the dynamical equations, relativistic effects considered, etc.  

The  formation of baryon-antibaryon states has also been studied within QCD sum rules~\cite{Wan:2021vny}. In case of the $\Lambda\bar\Lambda$ system, in Ref.~\cite{Wan:2021vny} a vector state with mass $2340\pm120$ MeV is obtained. Although the authors of Ref.~\cite{Wan:2021vny} concluded that the state found is above the $\Lambda\bar\Lambda$ threshold, the uncertainty determined for the mass of the state produces a lower limit of $2220$ MeV. Thus, the association of the $1^-$ $\Lambda\bar\Lambda$ state obtained in Ref.~\cite{Wan:2021vny} with $\phi(2170)$, although being less plausible, cannot be completely discarded with the present uncertainty in the experimental data~\cite{ParticleDataGroup:2022pth}.

\section{Conclusions}
In this work we have studied the decay modes of $\phi(2170)$ to $\phi\eta$ and $\phi\eta^\prime$. To do this, we consider $\phi(2170)$ to be a state generated from the dynamics involved in the $\phi f_0(980)$ system, with $f_0(980)$ being generated from the interaction of two pseudoscalars in the $s$-wave. Such internal structures for $\phi(2170)$ and $f_0(980)$ produces mechanisms for decaying to $\phi\eta$ and $\phi\eta^\prime$ involving triangular loops. The results obtained for the decay widths of $\phi(2170)$ to $\phi\eta$ and $\phi\eta^\prime$ indicate that the former is larger by a factor $R_{\eta/\eta^\prime}\simeq 2.6-5.2$. Comparing this answer with the one obtained using the BESIII and Belle results for the branching fractions of $\phi(2170)$, we find that our lower limit of $2.6$ is close to the upper limit obtained for $R_{\eta/\eta^\prime}$ by using the solutions II of the BESIII or III, IV of the Belle collaborations. We also find that models with different inner structures for $\phi(2170)$ may produce similar values for $R_{\eta/\eta^\prime}$, even if the predictions for other observables are different. Data with higher precision are necessary to understand the nature of $\phi(2170)$.

\section{Acknowledgments}
This work is supported by the Funda\c c\~ao de Amparo \`a Pesquisa do Estado de S\~ao Paulo (FAPESP), processos n${}^\circ$ 2023/01182-7, 2022/08347-9 and 2020/00676-8, and by the Conselho Nacional de Desenvolvimento Cient\'ifico e Tecnol\'ogico (CNPq), grant  n${}^\circ$ 305526/2019-7 and 303945/2019-2.

\appendix*
\section{Amplitudes for the pseudoscalar-pseudoscalar interaction}\label{f0ap}
In this appendix we provide the amplitudes for the processes $\mathcal{P}_i\bar{\mathcal{P}}^\prime_i\to \mathcal{P}_j\bar{\mathcal{P}}^\prime_j$ projected on total isospin 0 and in the $s$-wave, with $\mathcal{P}_i$, $\bar{\mathcal{P}}^\prime_i$ ($\mathcal{P}_j$, $\bar{\mathcal{P}}^\prime_j$) being the pseudoscalars constituting the channel $i$ ($j$).  In our approach, we consider the following coupled channels in the isospin base: $K\bar K$ (channel number 1), $\pi\pi$ (2), $\eta\eta$ (3), $\eta\eta^\prime$ (4), $\eta^\prime\eta^\prime$ (5). We also follow the isospin phase convention $|\pi^+\rangle=-|I=1,I_3=+1\rangle$, $|K^+\rangle=-|I=1/2,I_3=1/2\rangle$, where $I$ represents the isospin of the particle and $I_3$ its third projection~\cite{Oller:1997ti}. Using the Clebsch-Gordan coefficients, for total isospin $\mathcal{I}=0$ of the two-pseudoscalar system, we have the following combinations:
\begin{align}
|\pi\pi;\mathcal{I}=0,\mathcal{I}_3=0\rangle&=-\frac{1}{\sqrt{3}}\Big[|\pi^+\pi^-\rangle+|\pi^-\pi^+\rangle+|\pi^0\pi^0\rangle\Big],\nonumber
\end{align}
\begin{align}
|K\bar K; \mathcal{I}=0,\mathcal{I}_3=0\rangle&=-\frac{1}{\sqrt{2}}\Big[|K^+K^-\rangle+|K^0\bar K^0\rangle\Big],\nonumber
\end{align}
\begin{align}
|\eta\eta;\mathcal{I}=0,\mathcal{I}_3=0\rangle&=|\eta^0\eta^0\rangle,\nonumber
\end{align}
\begin{align}
|\eta\eta^\prime;\mathcal{I}=0,\mathcal{I}_3=0\rangle&=|\eta^0\eta^{\prime 0}\rangle,\nonumber
\end{align}
\begin{align}
|\eta^\prime\eta^\prime;\mathcal{I}=0,\mathcal{I}_3=0\rangle&=|\eta^{\prime 0}\eta^{\prime 0}.\rangle\label{isoscomb}
\end{align}
The isospin 0 projected amplitudes, $V_{ij}$, and loop functions, $G_i$, are then obtained as
\begin{align}
V_{ij}&=\langle \mathcal{P}_j\bar{\mathcal{P}}^\prime_j;I=0,I_3=0|V|\mathcal{P}_i\bar{\mathcal{P}}^\prime_i;I=0,I_3=0\rangle,\nonumber\\
G_i&=\langle \mathcal{P}_j\bar{\mathcal{P}}^\prime_j;I=0,I_3=0|G|\mathcal{P}_i\bar{\mathcal{P}}^\prime_i;I=0,I_3=0\rangle.\label{VIsos}
\end{align}
Equation (\ref{VIsos}) provides $V_{ij}$ and $G_i$ in terms of the corresponding ones in the charge basis. The amplitudes in the charge basis are obtained from the Lagrangian in Eq.~(\ref{LPP}) by using the Feynman rules. To calculate $G_i$, as well as $V_{ij}$, we consider an average mass for the members of the same isospin multiplet.

When solving Eq.~(\ref{BS}) in the center-of-mass frame of the system, the function $G_i/2$ needs to be used whenever $i$ is a channel constituted by two identical pseudoscalars. By doing this, double counting is avoided when integrating on $d^3q$ the terms in the series, obtained by iteration, of Eq.~(\ref{BS}). Such a procedure is equivalent to adding a factor $1/\sqrt{2}$ in those isospin states~\cite{Oller:1997ti} of Eq.~(\ref{isoscomb}) involving identical particles in the isospin basis. This latter normalization of the states is sometimes referred to as the unitary normalization~\cite{Oller:1997ti,Lin:2021isc}. The only difference is that in the latter case, the $t$-matrix element $t_{ij}$ obtained by solving Eq.~(\ref{BS}) needs to be multiplied by a factor $\sqrt{2}^{n_i+n_j}$, where $n_i$ ($n_j$) is 1 if channel $i$ ($j$) is constituted by two identical pseudoscalars, otherwise it is 0. 

In the following, we list the results obtained for $V_{ij}$ in terms of the $\eta-\eta^\prime$ mixing angle $\beta$. To simplify the notation, we define the factors $C_\beta\equiv \text{cos}\beta$ and $S_\beta\equiv\text{sin}\beta$, with $\beta$ being the mixing angle of Eq.~(\ref{beta}): 
\begin{align}
V_{11}&=-\frac{3s}{4 f^2_1},\label{V11}
\end{align}
\begin{align}
V_{12}&=-\frac{1}{2}\sqrt{\frac{3}{2}}\frac{s}{ f_1 f_2},
\end{align}
\begin{align}
&V_{13}=\frac{1}{6\sqrt{2} f_1f_3}\Bigg[8 S_\beta(\sqrt{2} C_\beta+S_\beta)m^2_K\nonumber\\
&\quad+C_\beta(9C_\beta s-2(C_\beta+2\sqrt{2}S_\beta)m^2_\pi-6 C_\beta m^2_\eta)\Bigg],
\end{align}
\begin{align}
&V_{14}=\frac{1}{6\sqrt{2} f_1f_4}\Bigg[-4(\sqrt{2}(C^2_\beta-S^2_\beta)+2 C_\beta S_\beta)m^2_K\nonumber\\
&\quad+2(\sqrt{2}(C^2_\beta-S^2_\beta)-C_\beta S_\beta)m^2_\pi\nonumber\\
&\quad+3 C_\beta S_\beta(3s-m^2_\eta-m^2_{\eta^\prime})\Bigg],
\end{align}
\begin{align}
&V_{15}=\frac{1}{6\sqrt{2} f_1 f_5}\Bigg[8 C_\beta(C_\beta-\sqrt{2}S_\beta)m^2_K\nonumber\\
&\quad+S_\beta(9s S_\beta+2(2\sqrt{2} C_\beta-S_\beta)m^2_\pi-6S_\beta m^2_{\eta^\prime})\Bigg]
\end{align}
\begin{align}
V_{22}&=\frac{-2s+m^2_\pi}{f^2_2},
\end{align}
\begin{align}
V_{23}&=\frac{(C^2_\beta+2 S^2_\beta-2\sqrt{2}C_\beta S_\beta)m^2_\pi}{\sqrt{3} f_2f_3}
\end{align}
\begin{align}
V_{24}&=\frac{(\sqrt{2} C_\beta-2S_\beta)(2C_\beta+\sqrt{2}S_\beta)m^2_\pi}{2\sqrt{3} f_2 f_4},\\
V_{25}&=\frac{(1+C^2_\beta+2\sqrt{2} C_\beta S_\beta)m^2_\pi}{\sqrt{3} f_2 f_5}
\end{align}
\begin{align}
&V_{33}=\frac{1}{9 f^2_3}\Bigg[-4(4 C^4_\beta+12 C^2_\beta S^2_\beta+8\sqrt{2} C^3_\beta S_\beta\nonumber\\
&\quad +4\sqrt{2} C_\beta S^3_\beta+S^4_\beta)m^2_K+(7C^4_\beta+12 C^2_\beta S^2_\beta\nonumber\\
&\quad+20\sqrt{2} C^3_\beta S_\beta+16\sqrt{2} C_\beta S^3_\beta-2S^4_\beta)m^2_\pi\Bigg],
\end{align}
\begin{align}
&V_{34}=-\frac{1}{324 f_3 f_4}\Bigg[(\sqrt{6}C_\beta-2\sqrt{3}S_\beta)^3(2\sqrt{3}C_\beta\nonumber\\
&\quad+\sqrt{6}S_\beta)m^2_\pi+24\sqrt{3}(\sqrt{2}C_\beta+S_\beta)^3\nonumber\\
&\quad\times(-\sqrt{3}C_\beta+\sqrt{6}S_\beta)(2m^2_K-m^2_\pi)\Bigg],
\end{align}
\begin{align}
&V_{35}=-\frac{1}{9 f_3 f_5}\Bigg[(2 C^4_\beta-3C^2_\beta S^2_\beta-2\sqrt{2}C^3_\beta S_\beta\nonumber\\
&\quad+2\sqrt{2}C_\beta S^3_\beta+2S^4_\beta)(4m^2_K-m^2_\pi)\Bigg],
\end{align}
\begin{align}
&V_{44}=-\frac{1}{9f^2_4}\Bigg[(2C^4_\beta-3C^2_\beta S^2_\beta-2\sqrt{2}C^3_\beta S_\beta\nonumber\\
&\quad+2\sqrt{2}C_\beta S^3_\beta+2S^4_\beta)(4m^2_k-m^2_\pi)\Bigg],
\end{align}
\begin{align}
&V_{45}=-\frac{1}{324f_4f_5}\Bigg[(\sqrt{6}C_\beta-2\sqrt{3}S_\beta)\nonumber\\
&\quad\times(2\sqrt{3}C_\beta+\sqrt{6}S_\beta)^3m^2_\pi\nonumber\\
&\quad+8\sqrt{3}(\sqrt{2}C_\beta+S_\beta)\nonumber\\
&\quad\times(-\sqrt{3}C_\beta+\sqrt{6}S_\beta)^3(2m^2_K-m^2_\pi)\Big],
\end{align}
\begin{align}
V_{55}&=-\frac{1}{9f^2_5}\Bigg[4(C^4_\beta+12C^2_\beta S^2_\beta-4\sqrt{2}C^3_\beta S_\beta\nonumber\\
&\quad-8\sqrt{2}C_\beta S^3_\beta+4S^4_\beta)m^2_K\nonumber\\
&\quad+(2C^4_\beta-12 C^2_\beta S^2_\beta+16\sqrt{2}C^3_\beta S_\beta\nonumber\\
&\quad+20\sqrt{2}C_\beta S^3_\beta-7S^4_\beta)m^2_\pi\Bigg].\label{V55}
\end{align}
Note that $V_{ij}=V_{ji}$, with $i,j=1,2,\dots 5$, and $f_i=\sqrt{f_{\mathcal{P}_i }f_{\bar{\mathcal{P^\prime}}_i}}$, with $\mathcal{P}_i$ and $\bar{\mathcal{P}}^\prime_i$ being the pseudoscalars constituting the channel $i$. 

When solving Eq.~(\ref{BS}), we have considered two cases: (I) different values for the pseudoscalar weak decay coupling constants in Eqs.~(\ref{V11})-(\ref{V55}), with $f_\pi=93$ MeV, $f_K=113$ MeV, $f_\eta=f_{\eta^\prime}=111$ MeV~\cite{Oller:1998hw}, and $Q_\text{max}=1020$ MeV in Eq.~(\ref{Gk}); (II) $f_\pi=f_K=f_\eta=f_{\eta^\prime}=93$ MeV and $Q_\text{max}=700$ MeV~\cite{Liang:2014tia,Xie:2014tma}. Note that different values of these pseudoscalar coupling constants are related to breaking of the SU(3) symmetry and the changes produced in $V$ can be reabsorbed in the model by readjusting the cutoff $Q_\text{max}$ while considering $f_\pi=f_K=f_\eta=f_\eta^\prime$.
\begin{figure}
\begin{tabular}{c}
\includegraphics[width=0.4\textwidth]{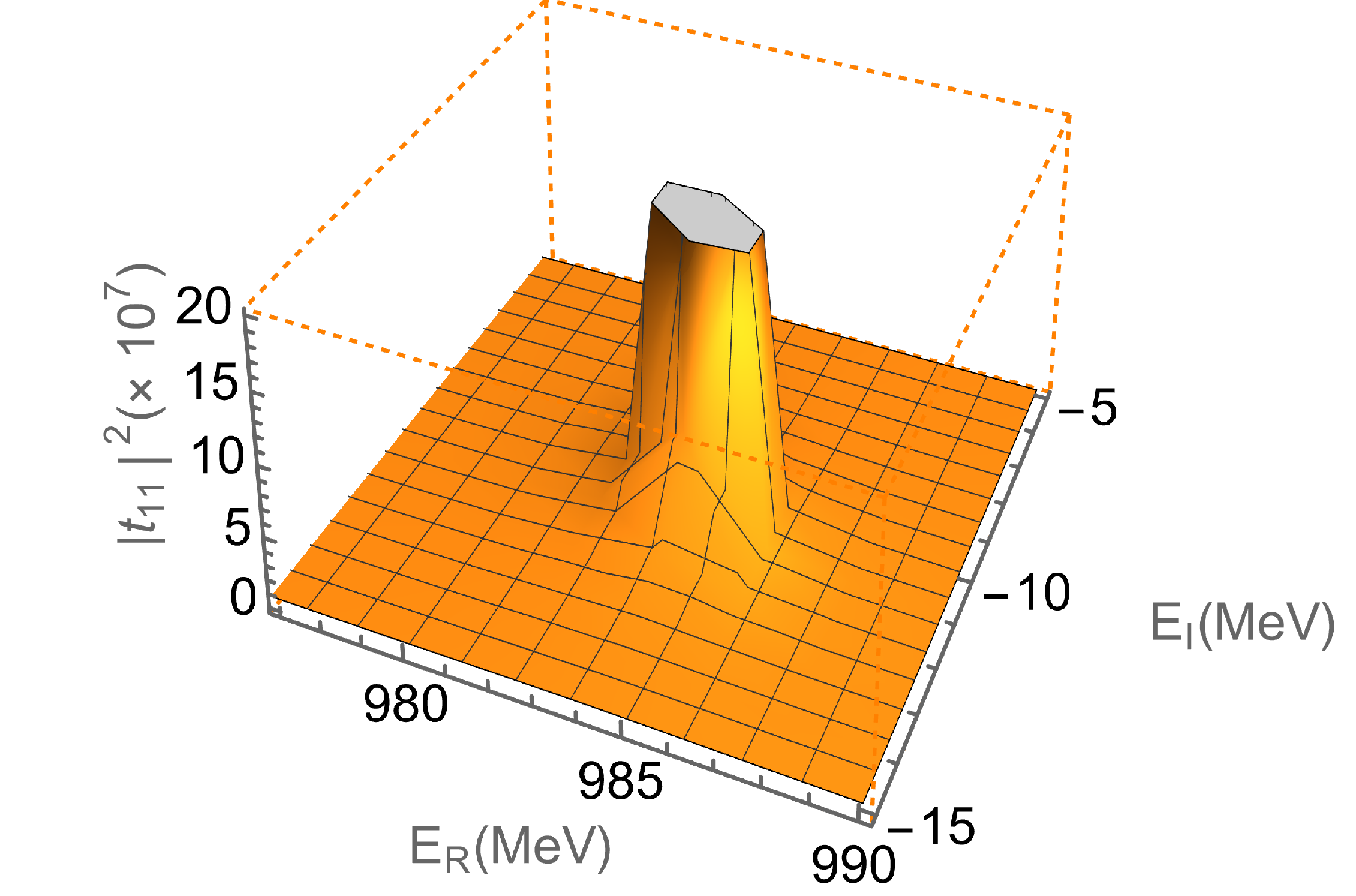}\\
\includegraphics[width=0.3\textwidth]{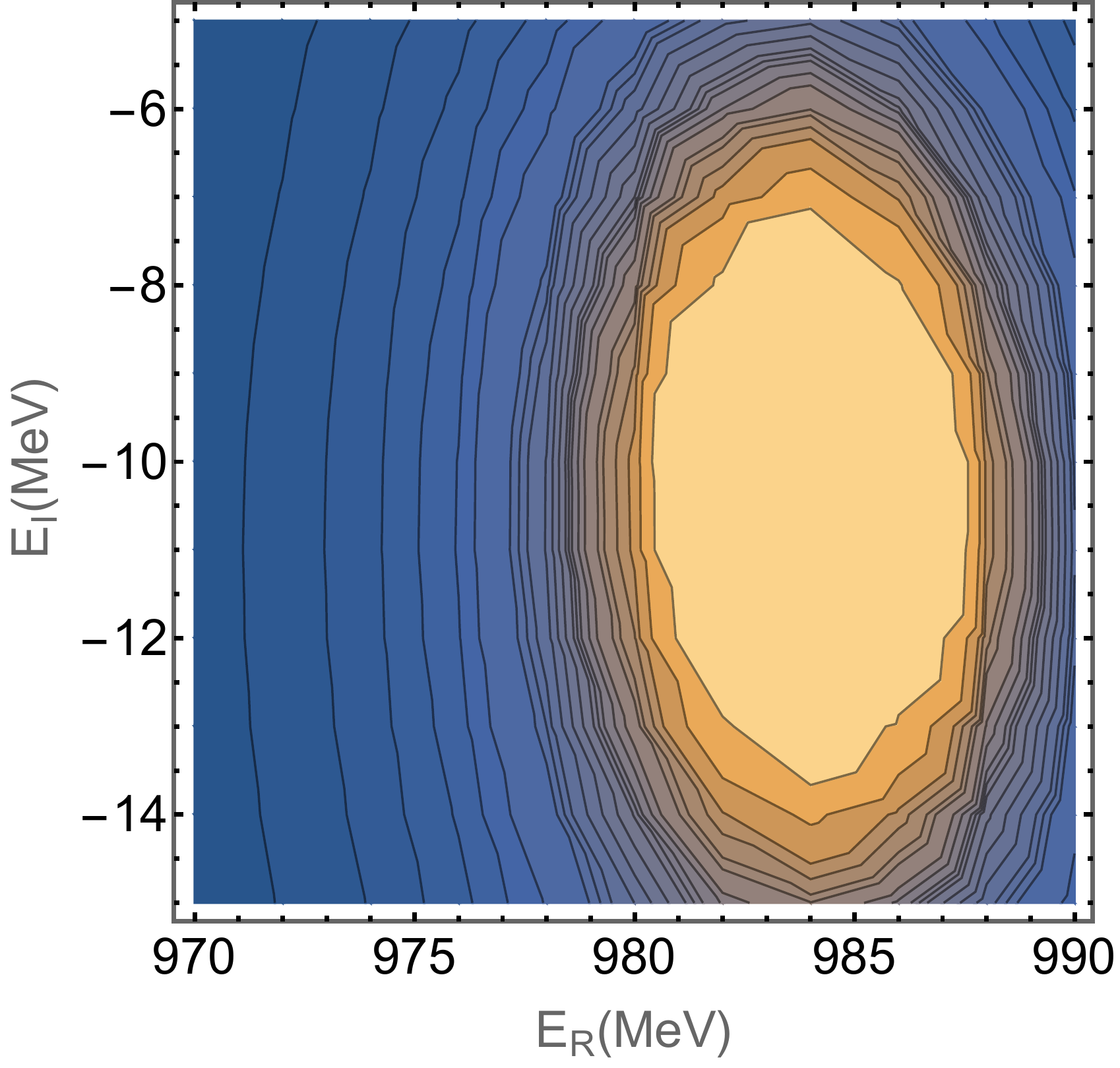}
\end{tabular}
\caption{(Top) Pole related to $f_0(980)$ generated as a consequence of the two-pseudoscalar dynamics in the $s$-wave. In the figure, the modulus squared of the $t$-matrix element for the transition $K\bar K\to K\bar K$ in isospin 0 is represented as a function of the real ($E_R$) and imaginary ($E_I$) parts of $\sqrt{s}$. Ideal $\eta-\eta^\prime$ mixing has been considered in this case. (Bottom) Contour plot for $|t_{11}|^2$.}\label{pole}
\end{figure}
\\\\
\begin{figure}[h!]
\centering
\includegraphics[width=0.45\textwidth]{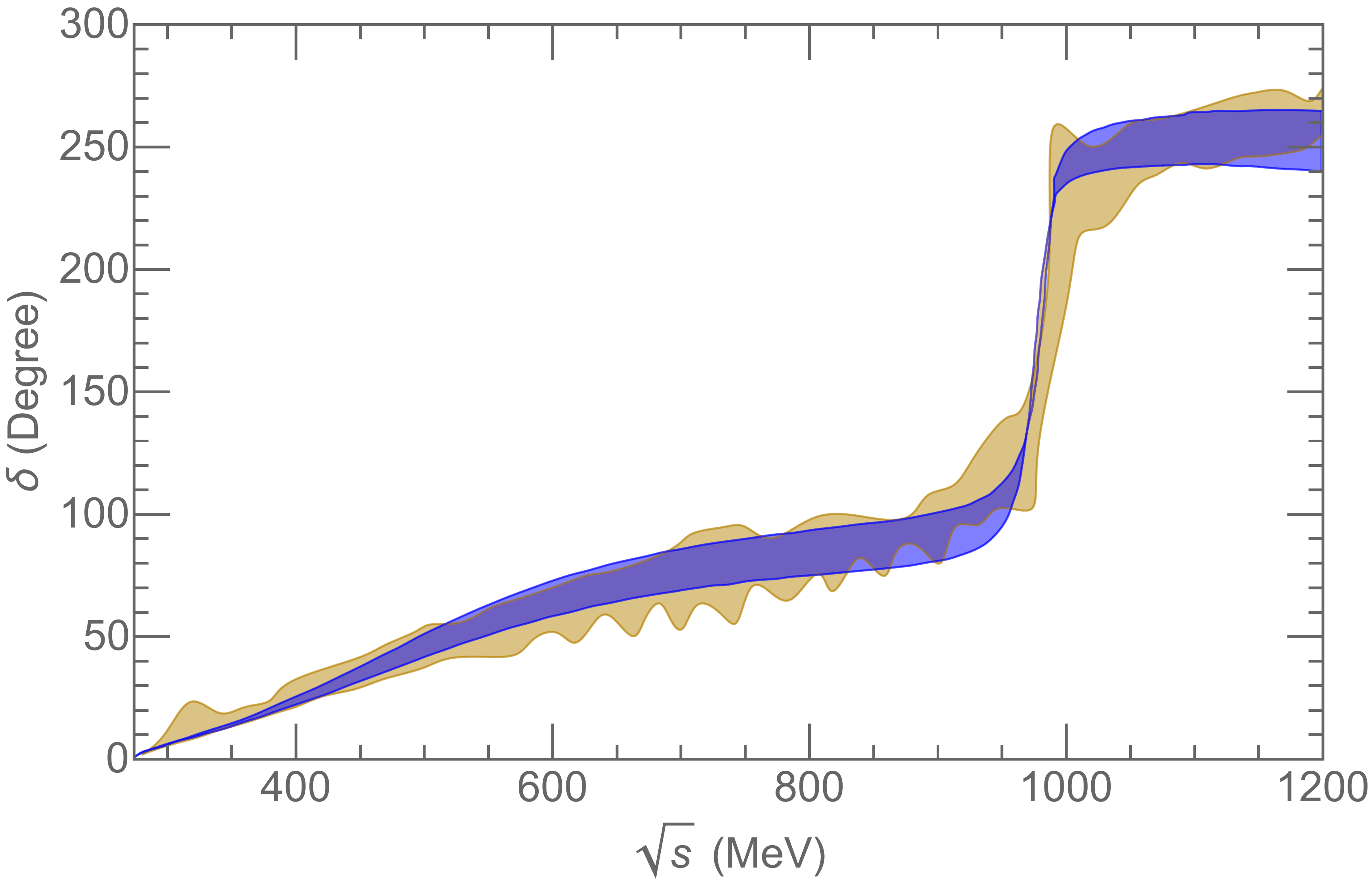}
\caption{Phase shift for $\pi\pi\to\pi\pi$ in the $s$-wave and isospin 0. The light shaded region corresponds to the boundaries obtained by considering the extremities of the error bars on the data~\cite{Hyams:1973zf,Estabrooks:1973dya,Grayer:1973at,Grayer:1974cr,Froggatt:1977hu}. The dark shaded region represents the uncertainty in our model.}\label{shift}
\end{figure}

Using the matrices
\begin{align}
V&=\left(\begin{array}{ccc}V_{11}&\cdots&V_{15}\\V_{21}&\cdots&V_{25}\\\vdots&\ddots&\vdots\\V_{51}&\cdots&V_{55}\end{array}\right),\nonumber\\
G&=\left(\begin{array}{cccc}G_{11}&0&\cdots&0\\0&G_{22}/2&\cdots&0\\\vdots&\vdots&\ddots&\vdots\\0&0&\cdots&G_{55}/2\end{array}\right),
\end{align}
we solve Eq.~(\ref{BS}) and search for poles of $t_{ij}$ in the complex energy plane. To do this the loop function needs to be analytically continued to the complex energy plane. This can be done by changing $G_i$ to $G_i-2i \text{Im}\{G_i\}$ whenever we are above the threshold of the channel $i$, with $G_i$ being determined in the first Riemann sheet~\cite{Oller:1997ti}.

In Fig.~\ref{pole} we show, for the case of ideal mixing, $|t_{11}|^2$ as a function of $E_R$ and $E_I$, with $E_R$ ($E_I$) being the real (imaginary) part of the complex energy $\sqrt{s}=E_R+i E_I$, and the corresponding contour plot. As shown in the figure, a pole at $\sqrt{s}_0=984.0-i 10.3$ MeV is obtained, characterizing an isospin 0 state with mass of 984 MeV and half-width of 10.3 MeV. This state can be related to $f_0(980)$, as can be seen in Fig.~\ref{shift}, where the $\pi\pi$ phase shift obtained in isospin 0 is represented and a good agreement with the experimental data is found.

Once the pole position is found and identified with $f_0(980)$, we can determine the couplings constants of $f_0$ to the channels considered to generate the state. To do this, we consider that close to the pole position, the matrix element $t_{ij}$ can be written as
\begin{align}
t_{ij}(s)=\frac{g_ig_j}{s-s_0},
\end{align}
where $g_i$ ($g_j$) is the coupling constant of $f_0(980)$ to the channel $i$ ($j$) and $s_0$ is the corresponding pole position (in the Mandelstam variable $s$). In this way, the product $g_i g_j$ can be interpreted as the residue of $t_{ij}$ at $s=s_0$. Considering a closed contour around the pole position $\sqrt{s}_0$ and Cauchy's theorem, we can determine the product $g_ig_j$ as
\begin{align}
g_ig_j=\frac{1}{2\pi i}\oint ds\, t_{ij}(s)=\frac{1}{\pi i}\oint dE E\, t_{ij}(E),\label{gigj}
\end{align}
with $E=\sqrt{s}$. Equation (\ref{gigj}) determines the couplings $g_i$, $i=1,2,\dots, 5$, up to a phase. In our case, we fix the phase by calculating $g_1$ by means of Eq.~(\ref{gigj}) and calculate the other couplings by using Eq.~(\ref{gigj}) for the product $g_1 g_j$, $j=2,3,\dots,5$. 

A remark about the $g_i$ couplings, obtained with the prescription $G_i\to G_i/2$  whenever channel $i$ is formed by two identical particles, is here in order. A process $f_0(980)\to \mathcal{P}\mathcal{P}$ can be described in terms of an effective Lagrangian given by
\begin{align}
\mathcal{L}=g\phi_{f_0}\phi_{\mathcal{P}}\phi_{\mathcal{P}},\label{Leff}
\end{align}
where $\phi_{f_0}$, $\phi_{\mathcal{P}}$ are, respectively, fields related to the $f_0(980)$ and the pseudoscalar $\mathcal{P}$ particles. In Eq.~(\ref{Leff}), $g$ is a coupling which reproduces the partial decay width of $f_0$ to the $\mathcal{P}\mathcal{P}$ channel, considering the latter to be an open channel for decay. The Lagrangian in Eq.~(\ref{Leff}) produces an amplitude for the process $f_0\to\mathcal{P}\mathcal{P}$ given by
\begin{align}
t_{f_0\to\mathcal{P}\mathcal{P}}=2g,\label{tf0PP}
\end{align}
where the factor 2 is a consequence of the different ways the two pseudoscalar fields in Eq.~(\ref{Leff}) can create the corresponding pseudoscalars particles in the final state. Using Eq.~(\ref{tf0PP}) and considering the nature (scalar and pseudoscalar) of the particles involved, the decay width of $f_0\to\mathcal{P}\mathcal{P}$ can be written as
\begin{align}
\Gamma_{f_0\to\mathcal{P}\mathcal{P}}=\frac{|\vec{p}_{\mathcal{P}}|}{16\pi m^2_{f_0}}|t_{f_0\to\mathcal{P}\mathcal{P}}|^2. \label{GammaA}
\end{align}
In Eq.~(\ref{GammaA}), $|\vec{p}_\mathcal{P}|$ is the modulus of the linear momentum of $\mathcal{P}$ in the rest frame of the decaying particle, $m_{f_0}$ is the mass of $f_0$ and a factor of $1/2!$ has been included as a consequence of the presence of two identical particles in the final state. Using Eq.~(\ref{tf0PP}), the preceding equation can be written as
\begin{align}
\Gamma_{f_0\to\mathcal{P}\mathcal{P}}=\frac{|\vec{p}_{\mathcal{P}}|}{8\pi m^2_{f_0}}(2|g|^2).\label{G1}
\end{align}
In terms of the coupling constant $g_{\mathcal{P}\mathcal{P}}$ determined by solving Eq.~(\ref{BS}) with the above mentioned prescription for the loop function, the partial decay width of $f_0\to\mathcal{P}\mathcal{P}$ can be obtained as
\begin{align}
\Gamma_{f_0\to\mathcal{P}\mathcal{P}}=\frac{|\vec{p}_{\mathcal{P}}|}{16\pi m^2_{f_0}}|g_{\mathcal{P}\mathcal{P}}|^2.\label{G2}
\end{align}
By comparing Eqs.~(\ref{G1}) and (\ref{G2}), we get that $|g_{\mathcal{P}\mathcal{P}}|=2|g|$. We can then conclude that using Eq.~(\ref{tf01}) already implements the factor $2$ present in Eq.~(\ref{tf0PP}).

\bibliographystyle{unsrt}
\bibliography{refs}

\begin{thebibliography}{10}

\bibitem{Okubo:1963fa}
S.~Okubo.
\newblock {Phi meson and unitary symmetry model}.
\newblock {\em Phys. Lett.}, 5:165--168, 1963.

\bibitem{Iizuka:1966fk}
Jugoro Iizuka.
\newblock {Systematics and phenomenology of meson family}.
\newblock {\em Prog. Theor. Phys. Suppl.}, 37:21--34, 1966.

\bibitem{BaBar:2006gsq}
Bernard Aubert et~al.
\newblock {A Structure at 2175-MeV in $e^{+} e^{-} \to \phi$ f0(980) Observed
  via Initial-State Radiation}.
\newblock {\em Phys. Rev. D}, 74:091103, 2006.

\bibitem{Belle:2008kuo}
C.~P. Shen et~al.
\newblock {Observation of the phi(1680) and the Y(2175) in e+e-
  ---\ensuremath{>} phi pi+ pi-}.
\newblock {\em Phys. Rev. D}, 80:031101, 2009.

\bibitem{BES:2007sqy}
Medina Ablikim et~al.
\newblock {Observation of Y(2175) in J / psi ---\ensuremath{>} eta phi
  f(0)(980)}.
\newblock {\em Phys. Rev. Lett.}, 100:102003, 2008.

\bibitem{Barnes:2002mu}
T.~Barnes, N.~Black, and P.~R. Page.
\newblock {Strong decays of strange quarkonia}.
\newblock {\em Phys. Rev. D}, 68:054014, 2003.

\bibitem{Ding:2007pc}
Gui-Jun Ding and Mu-Lin Yan.
\newblock {Y(2175): Distinguish Hybrid State from Higher Quarkonium}.
\newblock {\em Phys. Lett. B}, 657:49--54, 2007.

\bibitem{Pang:2019ttv}
Cheng-Qun Pang.
\newblock {Excited states of $\phi$ meson}.
\newblock {\em Phys. Rev. D}, 99(7):074015, 2019.

\bibitem{Li:2020xzs}
Qi~Li, Long-Cheng Gui, Ming-Sheng Liu, Qi-Fang L\"u, and Xian-Hui Zhong.
\newblock {Mass spectrum and strong decays of strangeonium in a constituent
  quark model}.
\newblock {\em Chin. Phys. C}, 45(2):023116, 2021.

\bibitem{Wang:2021gle}
Jun-Zhang Wang, Li-Ming Wang, Xiang Liu, and Takayuki Matsuki.
\newblock {Deciphering the light vector meson contribution to the cross
  sections of e+e- annihilations into the open-strange channels through a
  combined analysis}.
\newblock {\em Phys. Rev. D}, 104(5):054045, 2021.

\bibitem{Ding:2006ya}
Gui-Jun Ding and Mu-Lin Yan.
\newblock {A Candidate for 1-- strangeonium hybrid}.
\newblock {\em Phys. Lett. B}, 650:390--400, 2007.

\bibitem{Page:1998gz}
Philip~R. Page, Eric~S. Swanson, and Adam~P. Szczepaniak.
\newblock {Hybrid meson decay phenomenology}.
\newblock {\em Phys. Rev. D}, 59:034016, 1999.

\bibitem{Barnes:1995hc}
Ted Barnes, F.~E. Close, and E.~S. Swanson.
\newblock {Hybrid and conventional mesons in the flux tube model: Numerical
  studies and their phenomenological implications}.
\newblock {\em Phys. Rev. D}, 52:5242--5256, 1995.

\bibitem{Ma:2020bex}
Yunheng Ma, Ying Chen, Ming Gong, and Zhaofeng Liu.
\newblock {Strangeonium-like hybrids on the lattice}.
\newblock {\em Chin. Phys. C}, 45(1):013112, 2021.

\bibitem{Wang:2006ri}
Zhi-Gang Wang.
\newblock {Analysis of the Y(2175) as a tetraquark state with QCD sum rules}.
\newblock {\em Nucl. Phys. A}, 791:106--116, 2007.

\bibitem{Chen:2008ej}
Hua-Xing Chen, Xiang Liu, Atsushi Hosaka, and Shi-Lin Zhu.
\newblock {The Y(2175) State in the QCD Sum Rule}.
\newblock {\em Phys. Rev. D}, 78:034012, 2008.

\bibitem{Deng:2010zzd}
Chengrong Deng, Jialun Ping, Fan Wang, and T.~Goldman.
\newblock {Tetraquark state and multibody interaction}.
\newblock {\em Phys. Rev. D}, 82:074001, 2010.

\bibitem{Chen:2018kuu}
Hua-Xing Chen, Cheng-Ping Shen, and Shi-Lin Zhu.
\newblock {A possible partner state of the $Y(2175)$}.
\newblock {\em Phys. Rev. D}, 98(1):014011, 2018.

\bibitem{Ke:2018evd}
Hong-Wei Ke and Xue-Qian Li.
\newblock {Study of the strong decays of $\phi(2170)$ and the future charm-tau
  factory}.
\newblock {\em Phys. Rev. D}, 99(3):036014, 2019.

\bibitem{Wang:2019nln}
Zhi-Gang Wang.
\newblock {Light tetraquark state candidates}.
\newblock {\em Adv. High Energy Phys.}, 2020:6438730, 2020.

\bibitem{Liu:2020lpw}
Feng-Xiao Liu, Ming-Sheng Liu, Xian-Hui Zhong, and Qiang Zhao.
\newblock {Fully-strange tetraquark $ss\bar{s}\bar{s}$ spectrum and possible
  experimental evidence}.
\newblock {\em Phys. Rev. D}, 103(1):016016, 2021.

\bibitem{Coito:2009na}
Susana Coito, George Rupp, and Eef van Beveren.
\newblock {Multichannel calculation of excited vector phi resonances and the
  phi(2170)}.
\newblock {\em Phys. Rev. D}, 80:094011, 2009.

\bibitem{Drenska:2008gr}
N.~V. Drenska, R.~Faccini, and A.~D. Polosa.
\newblock {Higher Tetraquark Particles}.
\newblock {\em Phys. Lett. B}, 669:160--166, 2008.

\bibitem{Agaev:2019coa}
S.~S. Agaev, K.~Azizi, and H.~Sundu.
\newblock {Nature of the vector resonance $Y(2175)$}.
\newblock {\em Phys. Rev. D}, 101(7):074012, 2020.

\bibitem{MartinezTorres:2008gy}
A.~Martinez~Torres, K.~P. Khemchandani, L.~S. Geng, M.~Napsuciale, and E.~Oset.
\newblock {The X(2175) as a resonant state of the phi K anti-K system}.
\newblock {\em Phys. Rev. D}, 78:074031, 2008.

\bibitem{Zhao:2013ffn}
Lu~Zhao, Ning Li, Shi-Lin Zhu, and Bing-Song Zou.
\newblock {Meson-exchange model for the $\Lambda\bar{\Lambda}$ interaction}.
\newblock {\em Phys. Rev. D}, 87(5):054034, 2013.

\bibitem{Dong:2017rmg}
Yubing Dong, Amand Faessler, Thomas Gutsche, Qifang L\"u, and Valery~E.
  Lyubovitskij.
\newblock {Selected strong decays of $\eta(2225)$ and $\phi(2170)$ as $\Lambda
  \bar\Lambda$ bound states}.
\newblock {\em Phys. Rev. D}, 96(7):074027, 2017.

\bibitem{BESIII:2021bjn}
Medina Ablikim et~al.
\newblock {Study of the process $e^{+}e^{-}\rightarrow\phi\eta$ at
  center-of-mass energies between 2.00 and 3.08 GeV}.
\newblock {\em Phys. Rev. D}, 104(3):032007, 2021.

\bibitem{BESIII:2020gnc}
M.~Ablikim et~al.
\newblock {Observation of a structure in $e^{+}e^{-} \to \phi \eta^{\prime}$ at
  $\sqrt{s}$ from 2.05 to 3.08 GeV}.
\newblock {\em Phys. Rev. D}, 102(1):012008, 2020.

\bibitem{BaBar:2007ceh}
Bernard Aubert et~al.
\newblock {Measurements of $e^{+} e^{-} \to K^{+} K^{-} \eta$, $K^{+} K^{-}
  \pi^0$ and $K^0_{s} K^\pm \pi^\mp$ cross- sections using initial state
  radiation events}.
\newblock {\em Phys. Rev. D}, 77:092002, 2008.

\bibitem{Belle:2022fhh}
{Study of $e^{+}e^{-}\to\eta\phi$ via Initial State Radiation at Belle}.
\newblock {\em arXiv: 2209.00810 [hep-ex]}.

\bibitem{BESIII:2020vtu}
M.~Ablikim et~al.
\newblock {Observation of a Resonant Structure in $e^{+}e^{-} \to
  K^{+}K^{-}\pi^{0}\pi^{0}$}.
\newblock {\em Phys. Rev. Lett.}, 124(11):112001, 2020.

\bibitem{Malabarba:2020grf}
Brenda~B. Malabarba, Xiu-Lei Ren, K.~P. Khemchandani, and A.~Martinez~Torres.
\newblock {Partial decay widths of $\phi(2170)$ to kaonic resonances}.
\newblock {\em Phys. Rev. D}, 103(1):016018, 2021.

\bibitem{ParticleDataGroup:2022pth}
R.~L. Workman et~al.
\newblock {Review of Particle Physics}.
\newblock {\em PTEP}, 2022:083C01, 2022.

\bibitem{Oller:1997ti}
J.~A. Oller and E.~Oset.
\newblock {Chiral symmetry amplitudes in the S wave isoscalar and isovector
  channels and the $\sigma$, f$_0$(980), a$_0$(980) scalar mesons}.
\newblock {\em Nucl. Phys. A}, 620:438--456, 1997.
\newblock [Erratum: Nucl.Phys.A 652, 407--409 (1999)].

\bibitem{Oller:1998hw}
J.~A. Oller, E.~Oset, and J.~R. Pelaez.
\newblock {Meson meson interaction in a nonperturbative chiral approach}.
\newblock {\em Phys. Rev. D}, 59:074001, 1999.
\newblock [Erratum: Phys.Rev.D 60, 099906 (1999), Erratum: Phys.Rev.D 75,
  099903 (2007)].

\bibitem{Oller:1998zr}
J.~A. Oller and E.~Oset.
\newblock {N/D description of two meson amplitudes and chiral symmetry}.
\newblock {\em Phys. Rev. D}, 60:074023, 1999.

\bibitem{Herrera-Siklody:1996tqr}
P.~Herrera-Siklody, J.~I. Latorre, P.~Pascual, and J.~Taron.
\newblock {Chiral effective Lagrangian in the large N(c) limit: The Nonet
  case}.
\newblock {\em Nucl. Phys. B}, 497:345--386, 1997.

\bibitem{Kaiser:2000gs}
Roland Kaiser and H.~Leutwyler.
\newblock {Large N(c) in chiral perturbation theory}.
\newblock {\em Eur. Phys. J. C}, 17:623--649, 2000.

\bibitem{Gilman:1987ax}
Frederick~J. Gilman and Russel Kauffman.
\newblock {The eta Eta-prime Mixing Angle}.
\newblock {\em Phys. Rev. D}, 36:2761, 1987.
\newblock [Erratum: Phys.Rev.D 37, 3348 (1988)].

\bibitem{Akhoury:1987ed}
R.~Akhoury and J.~M. Frere.
\newblock {$\eta$, $\eta^\prime$ Mixing and Anomalies}.
\newblock {\em Phys. Lett. B}, 220:258--264, 1989.

\bibitem{Bramon:1997mf}
A.~Bramon, R.~Escribano, and M.~D. Scadron.
\newblock {Mixing of eta - eta-prime mesons in J / psi decays into a vector and
  a pseudoscalar meson}.
\newblock {\em Phys. Lett. B}, 403:339--343, 1997.

\bibitem{Venugopal:1998fq}
E.~P. Venugopal and Barry~R. Holstein.
\newblock {Chiral anomaly and eta - eta-prime mixing}.
\newblock {\em Phys. Rev. D}, 57:4397--4402, 1998.

\bibitem{Liang:2014tia}
W.~H. Liang and E.~Oset.
\newblock {$B^0$ and $B^0_s$ decays into $J/\psi$ $f_0(980)$ and $J/\psi$
  $f_0(500)$ and the nature of the scalar resonances}.
\newblock {\em Phys. Lett. B}, 737:70--74, 2014.

\bibitem{Lin:2021isc}
Jia-Xin Lin, Jia-Ting Li, Sheng-Juan Jiang, Wei-Hong Liang, and E.~Oset.
\newblock {The $D_s^+ \rightarrow a_0(980) e^+ \nu _e$ reaction and the
  $a_0(980)-f_0(980)$ mixing}.
\newblock {\em Eur. Phys. J. C}, 81(11):1017, 2021.

\bibitem{Xie:2014tma}
Ju-Jun Xie, Lian-Rong Dai, and Eulogio Oset.
\newblock {The low lying scalar resonances in the $D^0$ decays into $K^0_s$ and
  $f_0(500)$, $f_0(980)$, $a_0(980)$}.
\newblock {\em Phys. Lett. B}, 742:363--369, 2015.

\bibitem{Fujiwara:1984mp}
Takanori Fujiwara, Taichiro Kugo, Haruhiko Terao, Shozo Uehara, and Koichi
  Yamawaki.
\newblock {Nonabelian Anomaly and Vector Mesons as Dynamical Gauge Bosons of
  Hidden Local Symmetries}.
\newblock {\em Prog. Theor. Phys.}, 73:926, 1985.

\bibitem{Bando:1985rf}
Masako Bando, Taichiro Kugo, and Koichi Yamawaki.
\newblock {On the Vector Mesons as Dynamical Gauge Bosons of Hidden Local
  Symmetries}.
\newblock {\em Nucl. Phys. B}, 259:493, 1985.

\bibitem{Bramon:1994pq}
A.~Bramon, A.~Grau, and G.~Pancheri.
\newblock {Radiative vector meson decays in SU(3) broken effective chiral
  Lagrangians}.
\newblock {\em Phys. Lett. B}, 344:240--244, 1995.

\bibitem{Gamermann:2009uq}
D.~Gamermann, J.~Nieves, E.~Oset, and E.~Ruiz~Arriola.
\newblock {Couplings in coupled channels versus wave functions: application to
  the X(3872) resonance}.
\newblock {\em Phys. Rev. D}, 81:014029, 2010.

\bibitem{Godfrey:1985xj}
S.~Godfrey and Nathan Isgur.
\newblock {Mesons in a Relativized Quark Model with Chromodynamics}.
\newblock {\em Phys. Rev. D}, 32:189--231, 1985.

\bibitem{Jaffe:1976ih}
Robert~L. Jaffe.
\newblock {Multi-Quark Hadrons. 2. Methods}.
\newblock {\em Phys. Rev. D}, 15:281, 1977.

\bibitem{Machleidt:1987hj}
R.~Machleidt, K.~Holinde, and C.~Elster.
\newblock {The Bonn Meson Exchange Model for the Nucleon Nucleon Interaction}.
\newblock {\em Phys. Rept.}, 149:1--89, 1987.

\bibitem{Weinberg:1962hj}
Steven Weinberg.
\newblock {Elementary particle theory of composite particles}.
\newblock {\em Phys. Rev.}, 130:776--783, 1963.

\bibitem{Burakovsky:1997sd}
L.~Burakovsky and J.~Terrance Goldman.
\newblock {Gell-Mann-Okubo mass formula revisited}.
\newblock 8 1997.

\bibitem{Klingl:1997kf}
F.~Klingl, Norbert Kaiser, and W.~Weise.
\newblock {Current correlation functions, QCD sum rules and vector mesons in
  baryonic matter}.
\newblock {\em Nucl. Phys. A}, 624:527--563, 1997.

\bibitem{Nagahiro:2011fi}
H.~Nagahiro, S.~Hirenzaki, E.~Oset, and A.~Ramos.
\newblock {eta-prime nucleus optical potential and possible eta-prime bound
  states}.
\newblock {\em Phys. Lett. B}, 709:87--92, 2012.

\bibitem{Borasoy:1998pe}
B.~Borasoy.
\newblock {Baryon axial currents}.
\newblock {\em Phys. Rev. D}, 59:054021, 1999.

\bibitem{Zhu:2019ibc}
Jun-Tao Zhu, Yi~Liu, Dian-Yong Chen, Longyu Jiang, and Jun He.
\newblock {$X$(2239) and $\eta (2225)$ as hidden-strange molecular states from
  $\Lambda \bar \Lambda$ interaction}.
\newblock {\em Chin. Phys. C}, 44(12):123103, 2020.

\bibitem{BESIII:2018ldc}
M.~Ablikim et~al.
\newblock {Measurement of $e^{+} e^{-} \rightarrow K^{+} K^{-}$ cross section
  at $\sqrt{s} = 2.00 - 3.08$ GeV}.
\newblock {\em Phys. Rev. D}, 99(3):032001, 2019.

\bibitem{Wan:2021vny}
Bing-Dong Wan, Sheng-Qi Zhang, and Cong-Feng Qiao.
\newblock {Light baryonium spectrum}.
\newblock {\em Phys. Rev. D}, 105(1):014016, 2022.

\bibitem{Hyams:1973zf}
B.~Hyams et~al.
\newblock {$\pi\pi$ Phase Shift Analysis from 600-MeV to 1900-MeV}.
\newblock {\em Nucl. Phys. B}, 64:134--162, 1973.

\bibitem{Estabrooks:1973dya}
P.~Estabrooks et~al.
\newblock {$\pi \pi$ Phase Shift Analysis}.
\newblock {\em AIP Conf. Proc.}, 13:37--79, 1973.

\bibitem{Grayer:1973at}
G.~Grayer et~al.
\newblock {Coupled Channel Analysis in the $K\bar{K}$ Threshold Region}.
\newblock {\em AIP Conf. Proc.}, 13:117--134, 1973.

\bibitem{Grayer:1974cr}
G.~Grayer et~al.
\newblock {High Statistics Study of the Reaction pi- p --\ensuremath{>} pi- pi+
  n: Apparatus, Method of Analysis, and General Features of Results at
  17-GeV/c}.
\newblock {\em Nucl. Phys. B}, 75:189--245, 1974.

\bibitem{Froggatt:1977hu}
C.~D. Froggatt and J.~L. Petersen.
\newblock {Phase Shift Analysis of pi+ pi- Scattering Between 1.0-GeV and
  1.8-GeV Based on Fixed Momentum Transfer Analyticity. 2.}
\newblock {\em Nucl. Phys. B}, 129:89--110, 1977.

\end{thebibliography}

\end{document}